\documentclass[aps,pra,superscriptaddress,twocolumn]{revtex4}
\usepackage{graphicx,amsmath,amssymb,amsfonts,latexsym,color,dcolumn,bm}

\newcommand{\beq}{\begin{equation}}
\newcommand{\eeq}{\end{equation}}
\newcommand{\bea}{\begin{eqnarray}}
\newcommand{\eea}{\end{eqnarray}}
\providecommand{\abs}[1]{\left\lvert#1\right\rvert}

\providecommand{\bra}[1]{\langle #1 \rvert}
\providecommand{\ket}[1]{\lvert #1 \rangle}

\usepackage{dsfont}
\usepackage{physics}
\usepackage{tocvsec2}
\usepackage[toc,page]{appendix}
\usepackage{physics}
\usepackage{tabularx}

\usepackage{tikz}

\begin{document}

\title{Generation of time-frequency grid state with integrated biphoton frequency combs }
\author{N. Fabre}\email{nicolas.fabre@univ-paris-diderot.fr}
\affiliation{Laboratoire Mat\'eriaux et Ph\'enom\`enes Quantiques, Sorbonne Paris Cit\'e, Universit\'e de Paris, CNRS UMR 7162, 75013 Paris, France}
\author{G. Maltese}
\affiliation{Laboratoire Mat\'eriaux et Ph\'enom\`enes Quantiques, Sorbonne Paris Cit\'e, Universit\'e de Paris, CNRS UMR 7162, 75013 Paris, France}
\author{F. Appas}
\affiliation{Laboratoire Mat\'eriaux et Ph\'enom\`enes Quantiques, Sorbonne Paris Cit\'e, Universit\'e de Paris, CNRS UMR 7162, 75013 Paris, France}

\author{S. Felicetti}
\affiliation{Laboratoire Mat\'eriaux et Ph\'enom\`enes Quantiques, Sorbonne Paris Cit\'e, Universit\'e de Paris, CNRS UMR 7162, 75013 Paris, France}

\author{A. Ketterer}
\affiliation{Laboratoire Mat\'eriaux et Ph\'enom\`enes Quantiques, Sorbonne Paris Cit\'e, Universit\'e de Paris, CNRS UMR 7162, 75013 Paris, France}
\affiliation{Physikalisches Institut, Albert-Ludwigs-Universit\"at Freiburg, Hermann-Herder-Str. 3, 79104 Freiburg, Germany}
\author{A. Keller}
\affiliation{Laboratoire Mat\'eriaux et Ph\'enom\`enes Quantiques, Sorbonne Paris Cit\'e, Universit\'e de Paris, Univ. Paris Saclay, CNRS UMR 7162, 75013 Paris, France}

\author{T. Coudreau}
\affiliation{Laboratoire Mat\'eriaux et Ph\'enom\`enes Quantiques, Sorbonne Paris Cit\'e, Universit\'e de Paris, CNRS UMR 7162, 75013 Paris, France}

\author{F. Baboux}
\affiliation{Laboratoire Mat\'eriaux et Ph\'enom\`enes Quantiques, Sorbonne Paris Cit\'e, Universit\'e de Paris, CNRS UMR 7162, 75013 Paris, France}

\author{M.I. Amanti}
\affiliation{Laboratoire Mat\'eriaux et Ph\'enom\`enes Quantiques, Sorbonne Paris Cit\'e, Universit\'e de Paris, CNRS UMR 7162, 75013 Paris, France}

\author{S. Ducci}
\affiliation{Laboratoire Mat\'eriaux et Ph\'enom\`enes Quantiques, Sorbonne Paris Cit\'e, Universit\'e de Paris, CNRS UMR 7162, 75013 Paris, France}

\author{P. Milman}
\affiliation{Laboratoire Mat\'eriaux et Ph\'enom\`enes Quantiques, Sorbonne Paris Cit\'e, Universit\'e de Paris, CNRS UMR 7162, 75013 Paris, France}

\date{\today}
 
\begin{abstract}
Encoding quantum information  in continuous variables is intrinsically faulty. Nevertheless, redundant qubits can be used for error correction, as proposed by Gottesman, Kitaev and Preskill in Phys. Rev. A \textbf{64} 012310, (2001).  We show how to experimentally implement this encoding using time-frequency continuous degrees of freedom of photon pairs produced by spontaneous parametric down conversion. We experimentally illustrate our results using an integrated AlGaAs photon pairs source. We show how single qubit gates can be implemented and finally propose a theoretical scheme for correcting errors in a circuit-like and in a measurement-based architecture. 
\end{abstract}

\vskip2pc 
 
\maketitle

\section {Introduction} 

Quantum information can be encoded in qubits corresponding to discrete quantum states of physical systems, such as atomic electronic states or the polarization of single photons. The essence of quantum computation (QC) is to manipulate qubits with a universal set of unitary quantum gates \cite{Braunstein, Nielsen}. A fundamental ingredient for QC, inherited from classical computation, is error correction. In the realm of quantum computing, quantum error correction (QEC) \cite{Gottesman,Steane} fights against a fundamental aspect of quantum systems: their fragility to keep quantum properties at large scale and for a long time that, usually, depends on the size of the system. Ingenious solutions to this problem consist in encoding a qubit of information in particular states composed of more than one physical qubit. The resulting logical qubits enabling QEC depend on the type of errors that are more likely to affect the system. For instance, a code that corrects for qubit flips, dephasing and all the errors composed by the combination of these ones involves the creation of a complex 5-qubit entangled state whose symmetries enable the detection and correction of the mentioned errors \cite{GottesThesis}. 

If one considers harmonic oscillators or analogous systems, as for instance two quadratures of the electromagnetic field, encoding quantum information in continuous variable (CV) is in principle possible using any two orthogonal states of a given quadrature \cite{Gaussian}. However, such states are non-physical, and the closest one can get to them is by considering squeezed states \cite{Squeezed}, which are sub shot-noise states. As a matter of fact, squeezed states can be considered as noisy quadrature eigenstates, where noise is modeled by a product between displacements in phase space and a Gaussian function representing the amplitude of probability distribution of displacements. Physical states described by continuous variables are thus intrinsically noisy and, within this picture, displacements in phase space are the main source of noise for such encoding. Moreover, since all physical states are noisy, errors propagate throughout quantum operations and must be corrected regularly. This picture is particularly suitable to a number of relevant physical systems, as the quantum state produced by optical parametric oscillators (OPO) \cite{Nicolas, Pfister, Furusawa} and continuous degrees of freedom of photon pairs, as it is discussed in the present manuscript.
The problem of correction from displacement errors was considered by Gottesman, Kitaev and Preskill (GKP), who introduced what we refer now on as GKP states \cite{GKP}, which are qubits defined in CV displaying a periodic structure.

In spite of the importance of GKP states in quantum information with CV  \cite{GKPTH,GKPTH1,GKPTH2}, their experimental engineering remains extremely difficult in quantum optics \cite{GKP1,GKP2,GKP3}. They correspond to highly non-Gaussian states composed of the coherent superposition of several delocalized states. The engineering of non-Gaussian states using OPOs is still challenging, and even if some experiments have demonstrated it \cite{Grangier, Trepssubstract, Laurat} they involve single photons addition and/or subtraction through post-selection. As a consequence, one of the major advantages of using such systems, determinacy, is lost. Also, the generated non-Gaussian states are still far from the physical GKP states. Recently, such states have been produced using the motional states of one trapped ion \cite{Ions}, in superconducting qubits \cite{supra}, and for others platforms \cite{GKPP3,GKP4,GKP5} have been developed. \\

In the present manuscript, we show that biphoton frequency combs produced by intracavity Spontaneous Parametric Down Conversion (SPDC) can be used to experimentally generate, manipulate and detect grid states encoded in time ($t$) and frequency ($\omega$). We propose a method to implement a fundamental operation of quantum error correction. Our results rely on the analogy between quantum states composed of many photons in one mode of the electromagnetic field and one photon that can occupy a continuum of frequency modes, which is the continuous degree of freedom we consider here. We show that the time-frequency phase space at the single photon level has the same non-commutative structure than the usual quadrature position-momentum phase space.    Our results are experimentally illustrated using an AlGaAs nonlinear cavity producing photon pairs by SPDC at room temperature and telecom wavelengths and compliant with electrical injection \cite{Boitier}.
  The proposed scheme is not specific to the considered platform and it could be implemented with other quantum-optical setups, such as those in \cite{Wang, Lu}.

This paper is organised as follows. We start in Sec. \ref{firstsection} by describing the properties of the time-frequency phase space of a single photon and describe its quantum structure and properties. We use these results to properly introduce the Chronocyclic Wigner distribution. In Sec. \ref{sectiontwo}, we define the time-frequency GKP states, or more generally the time-frequency grid states \cite{gridstate}, which are fully analogous to the ones made from position-momentum quadrature states \cite{GKP}. Following the formalism developed, we mathematically introduce the 2D time-frequency GKP state. In Sec. \ref{fourA}, we present an experimental implementation of our results in a chip-integrated source consisting of a AlGaAs Bragg reflector waveguide generating a 2D time-frequency grid state and in Sec. \ref{fourB}  how such states can be used for correcting  time-frequency shift errors. This state can be experimentally manipulated and characterised using HOM interferometry, as shown in Sec. \ref{fourC}. Finally in Sec. \ref{sectionMBQC}, we theoretically propose a simple experimental  scheme based on photon detection to implement quantum error correction for time-frequency GKP states.

\section{Time-frequency phase space of a single photon}\label{firstsection}

In this section, we introduce the time-frequency phase space of a single photon, following the construction of the usual non-relativistic phase space in quadrature position-momentum variable.

\subsection{Time-frequency phase space description}
In order to fix the notation, the photon creation operator at frequency $\omega$ acts on the vacuum $\ket{0}$ as:
\begin{equation}
\hat{a}^{\dagger}(\omega)\ket{0}=\ket{1_{\omega}}=\ket{\omega},
\end{equation}
 We thus  have  that $\ket{1_{\omega}}$ and $\ket{\omega}$ are equivalent notations for a single photon at frequency $\omega$. In the present manuscript,  we will be mostly interested in the single photon subspace  we will use the simplified notation $\ket{\omega}$. We can also define an annihilation operator fore the mode $\omega$ as $\hat{a}(\omega)\ket{\omega}=\ket{0}$.
If we consider only the frequency degree of freedom, the creation and annihilation operators obey the bosonic commutation relation:
\begin{equation}\label{bosonicop}
[\hat{a}(\omega),\hat{a}^{\dagger}(\omega')]=\delta(\omega-\omega')\mathds{I},
\end{equation}
where $\mathds{I}$ is the identity operator.
Analogously, we define the creation operator for a single photon at time $t$, $\hat{a}^{\dagger}(t)$,  where $t$ is the time interval elapsed from the photon's creation at the source and its arrival at the detector. The creation operator $\hat{a}^{\dagger}(t)$ can be obtained from a Fourier transform of  $\hat{a}^{\dagger}(\omega)$:
\begin{equation}
\hat{a}^{\dagger}(t)=\frac{1}{\sqrt{2\pi}}\int_{ \mathbb{R}} \text{d}\omega e^{i\omega t} \hat{a}^{\dagger}(\omega).
\end{equation}
And we have that: $\ket{t}=\ket{1_{t}}=\hat{a}^{\dagger}(t)\ket{0}=\frac{1}{\sqrt{2\pi}}\int_{ \mathbb{R}} \text{d}\omega e^{i\omega t} \ket{\omega}$.

Since $\{\ket{\omega}\}$ is an orthogonal basis, we can expand a pure single photon state $ \ket{\Psi}$ in this basis:
\begin{equation}\label{singlefrequency}
\ket{\Psi}=\int_{ \mathbb{R}}  S(\omega) \text{d}\omega \ket{\omega},
\end{equation}
where $S(\omega)$ is the amplitude spectrum of the single photon, with the normalization condition $\int_{\mathds{R}}  \abs{S(\omega)}^{2} \text{d}\omega =1$.

We also consider single photons with a temporal structure, described by the state:
\begin{equation}
\ket{\Psi}=\int_{ \mathbb{R}} \tilde{S}(t) \text{d}t \ket{t}.
\end{equation}
where $\tilde{S}(t)$ is the Fourier transform of the amplitude spectrum $S(\omega)$ of the source.
Free space propagation for a time $t$ leads to the evolution of the creation operator: $\hat{a}^{\dagger}_{t}(\omega)=e^{-i\omega t} \hat{a}^{\dagger}(\omega)$, and the wave function at time $t$ reads:
\begin{equation}\label{single}
\ket{\Psi(t)}=\int_{\mathds{R}} S(\omega)e^{-i\omega t} \text{d}\omega\ket{\omega}. 
\end{equation}
Rigorously, the integration range in Eq.~\eqref{single} and Eq.~\eqref{singlefrequency} should be $\mathbb{R}^{+}$. We have extended
it to  $\mathbb{R}$ as we consider experiments where the amplitude spectrum fulfills  $S(\omega \leq 0)=0$. In the experimental setup we implement in this paper, $S(\omega)$ is typically non-zero only in the telecom wavelength.

We now define the time-frequency Wigner distribution. The Wigner distribution in phase space can be seen as the expectation value of the parity operator \cite{ParityWigner} or, equivalently, as the inverse Fourier transform of a characteristic function. The latter is constructed using  a symmetric ordering of bosonic operators.
We will proceed analogously, by introducing the displacement frequency mode operator in the single photon subspace.
 
\subsubsection{Frequency-time Wigner distribution for a single photon}

Using the previously introduced bosonic operators, we can define the displacement mode operator in frequency as: 
\begin{equation}\hat D(\mu)=\int \hat a^{\dagger}(\omega+\mu)\hat a(\omega) \text{d}\omega,\end{equation}
Analogously, for the displacement in time we can write
\begin{equation}\hat {\cal{D}}(\tau)=\int \hat a^{\dagger}(t+\tau)\hat a(t) \text{d}t,\end{equation}
As for Eq.~\eqref{single}, we have extended the range of integration from 
$\mathbb{R}^+$ to $\mathbb{R}$. Indeed,  for the envisaged application, $\mu$ will be small enough so that the resulting displaced state amplitude spectrum $S(\omega)$ will have its support in $\mathbb{R}^+$. We have that $\hat D(\mu)\ket{\omega}=\ket{\omega+\mu}$ and $\hat D(\mu)\ket{t}=e^{i\mu t}\ket{t}$.

As in the usual phase space case, displacement operators do not commute, and we obtain the Weyl relation,
\begin{equation}\label{Weyl}
\hat D(\mu)\hat {\cal{D}}(\tau)=e^{i\mu \tau} \hat {\cal{D}}(\tau)\hat D(\mu).
\end{equation} 
Using the commutation relations, in analogy to the quadrature position-momentum phase space case, we can identify different possible orderings of the operators:  the normal order $\hat D_{n}(\mu,\tau)=\hat D(\mu)\hat {\cal{D}}(\tau)$, the anti-normal order $\hat D_{an}(\mu,\tau)=\hat {\cal{D}}(\tau)\hat D(\mu)$ and the symmetric order, $\hat D_{s}(\mu,\tau)= \hat D(\mu)\hat {\cal{D}}(\tau)e^{-i\tau\mu/2}$. The unitary displacement operators $\hat  D_{\xi}$, irrespectively of the ordering, $\xi=s, an, n$, obey the following orthogonality relation:

\begin{equation}\label{complete}\text{Tr}[\hat D_{\xi}^{\dagger}(\mu,\tau)\hat D_{\xi}(\mu',\tau')]=\delta(\tau'-\tau)\delta(\mu'-\mu), \end{equation}
and the completeness relation:
\begin{equation}\label{complete1} \iint \text{d}\mu\text{d}\tau \hat{D}_{\xi}(\mu,\tau) \hat{D}^{\dagger}_{\xi}(\mu,\tau) =\mathds{I}. \end{equation}
Using Eq. (\ref{complete}) and Eq. (\ref{complete1}) we can expand all Hermitian matrices in this orthogonal basis, thus the density matrix reads:

\begin{equation}\label{wignertorho}
 \hat \rho=\iint \chi_{\rho, \xi}(\mu,\tau)\hat D_{\xi}(\mu,\tau) \text{d}\mu \text{d}\tau. \end{equation}
The coordinate function $\chi_{\rho, \xi}(\mu,\tau)=\text{Tr}(\hat \rho \hat D_{\xi}^{\dagger}(\mu,\tau))$ is called the characteristic function, and it can be normal, anti-normal or symmetric depending on the ordering of the displacement operator.
The Fourier transform of the characteristic function leads to a quasi-probability distribution. In particular, using the symmetric characteristic distribution, one can obtain the chronocyclic Wigner distribution, 
\begin{equation}\label{Wigner}
W(\omega,t)=\frac{1}{\sqrt{2\pi}} \int \text{d}\omega' e^{2i\omega't} \bra{\omega-\omega'}\hat{\rho}\ket{\omega+\omega'}.
\end{equation}
The chronocyclic Wigner distribution gives the same information than the associated density matrix, following the completeness property. This distribution is also normalised: $\iint \text{d}t\text{d}\omega W(\omega,t)=\text{Tr}(\hat{\rho})=1$.
In the case of a pure state $\hat{\rho}=\ket{\psi}\bra{\psi}$ characterised by its amplitude spectrum $S(\omega)$ Eq.~\eqref{singlefrequency}, the chronocylic Wigner distribution can be written as:
\begin{equation}\label{rhotowigner}
W(\omega,t)=\frac{1}{\sqrt{2\pi}} \int \text{d}\omega' e^{2i\omega't} S(\omega-\omega')S^{*}(\omega+\omega').
\end{equation}
The marginals of the  Wigner distribution lead to different physically measurable quantities such as the spectrum of the source:
\begin{equation}\label{marginal1}
\int W(\omega,t) \text{d}t =\abs{S(\omega)}^{2},
\end{equation}
and the distribution of the arrival time of the photon of the source (using a fixed origin of time):
\begin{equation}\label{marginal2}
\int W(\omega,t) \text{d}\omega=\abs{\tilde{S}(t)}^{2} .
\end{equation}
We can also see the  chronocyclic Wigner distribution here as the expectation value of the displaced parity operator by applying the same methods as in \cite{ParityWigner} using the symmetric displacement operator $\hat D_{s}$:
\begin{equation}
W(\omega,t)=\text{Tr}(\hat{\rho}\hat{D}_{s}(\omega,t)\hat{\Pi}\hat{D}^{\dagger}_{s}(\omega,t)),
\end{equation}
where $\hat{\Pi}$ is the parity operator which acts on a frequency state as $\hat{\Pi}\ket{\omega}=\ket{-\omega}$. Consequently, measuring the chronocyclic Wigner distribution at the origin is a measurement of the average value of the parity operator. \\

Finally, it can be shown that the chronocyclic Wigner distribution obeys the Stratonovich-Weyl rules \cite{Strato1,Strato2}.

Owing to the non-commutativity of the bosonic operator Eq. (\ref{bosonicop}) and, consequently, the non-commutativity of the displacement operators (see Eq.~(\ref{Weyl})), the time-frequency phase space of a single photon is non-commutative, as the quadrature position-momentum phase space. It leads to the analogy between one single photon in many frequency modes and many photons in one frequency mode.

The introduced Wigner distribution can be generalized to the situation where more than one  photon occupy different frequency modes. We will describe in details the two photon case in  the next section.

\subsubsection{Wigner distribution of two photons and associated marginals}\label{wignertwo}

For a two photon state, the wave function can be written as:

\begin{equation}\ket{\psi}=\iint \text{d}\omega_{s} \text{d}\omega_{i}\text{JSA}(\omega_{s},\omega_{i})\ket{\omega_{s},\omega_{i}} ,\end{equation}
where the $\text{JSA}$ denotes the Joint Spectral Amplitude and $\omega_{s} (\omega_{i})$ is the frequency of the signal (idler) photon. 
The Wigner distribution of a pure state $\ket{\psi}$ is:
\begin{multline}  W(\omega_{s},\omega_{i},t_{s},t_{i})=\iint \text{d}\omega'\text{d}\omega'' e^{2i\omega't_{s}}e^{2i\omega''t_{i}}\\\bra{\omega_{s}-\omega',\omega_{i}-\omega''}\ket{\psi}\bra{\psi}\ket{\omega_{s}+\omega',\omega_{i}+\omega''},\end{multline}
with marginals:
\begin{equation}
\iint W(\omega_{s},\omega_{i},t_{s},t_{i}) \text{d}t_{s} \text{d}t_{i} =\text{JSI}(\omega_{s},\omega_{i}),
\end{equation}
where $\text{JSI}(\omega_{s},\omega_{i})=\abs{\text{JSA}(\omega_{s},\omega_{i})}^{2}$ is the Joint Spectral Intensity. We also have
\begin{equation}
\iint W(\omega_{s},\omega_{i},t_{s},t_{i}) \text{d}\omega_{s} \text{d}\omega_{i} =\text{JTI}(t_{s},t_{i}),
\end{equation}
where $\text{JTI}$ is the Joint Temporal Intensity, which is the probability to measure a photon at an arrival time $t_{s}$ in one detector and a photon at an arrival time $t_{i}$ in another detector.
We can also define two other ``crossed" marginals: the probability to detect one photon at the arrival time $t_{s}$ (resp. $t_{i}$) and the other at the frequency $\omega_{i}$ (resp. $\omega_{s}$). The measurement of the four marginals and the reconstruction of the JSA has been performed in \cite{TomographyJSA}, however this technique cannot be applied to all optical systems and depend on the spectral width of the considered JSA.

In the next section, we use the formalism presented here to analyse the phase space distribution and representation for some relevant quantum states. As a first example, we will describe coherent-like state in time-frequency variables. Then, we will address grid and GKP state formed with single and two-photons.

\section{Time-frequency grid states}\label{sectiontwo}

Before introducing grid states, we describe coherent-like states in time-frequency phase space.

\subsection{Coherent-like state in time-frequency variables}

We define the coherent-like state $\ket{\alpha}$ in time-frequency variable as a single photon state with a Gaussian amplitude spectrum:
\begin{equation}
\ket{\alpha}=\int \text{d}\omega e^{-(\omega-\omega_{1})^{2}/(2\Delta\omega^{2})} e^{i\omega \tau} \ket{\omega}
\end{equation}
where $\text{Re}(\alpha)=\frac{\omega_{1}}{\Delta\omega}$ and $\text{Im}(\alpha)=\tau$ are two parameters representing the average value of the chronocyclic Wigner distribution and $\Delta\omega$ is the spectral width of the Gaussian spectrum. As in the usual phase-space representation, the coherent-like state also forms an overcomplete basis.

\subsection{Grid states}

We now use the formalism developed in the previous section to introduce the time-frequency grid state and, more specifically, the time-frequency GKP state. We also show their application for time-frequency quantum error correction.\\

Grid states were first defined using quadrature position-momentum continuous variables \cite{GKP, gridsensor} and correspond to a two dimensional lattice in phase space where the area of the unit cell is a multiple of $2\pi$. Time-frequency grid states also verify the following property: they are periodic structures formed by the superposition of many modes of single photon. Such states are mathematically equivalent to the quadrature position-momentum ones due to the non-commutative algebra of the time-frequency displacement operators.

More precisely, the time-frequency grid state is defined as a common eigenstate with the eigenvalue $+1$ of the two commuting operators $\hat{D}(\omega)$ and ${\cal{\hat{D}}}(t)$  when the product of the two parameters verify $\omega t =0 \ \text{mod} \ 2\pi$. This property allows to measure both the time and the frequency modulo $2\pi$. The wave function of a general grid state can be written as:
\begin{equation}
\ket{\psi}=\sum_{n\in\mathds{Z}} \ket{n\alpha\sqrt{2\pi}}_{\omega},
\end{equation}
where $\alpha$ is an integer. The +1 eigenspace of the displacement operators is two dimensional and allows to define a qubit. It is developed in more details in the following paragraph.

\subsection{Frequency-time GKP state}

\subsubsection{Definition and notation}

\begin{figure}[h!]
\includegraphics[width=0.5\textwidth]{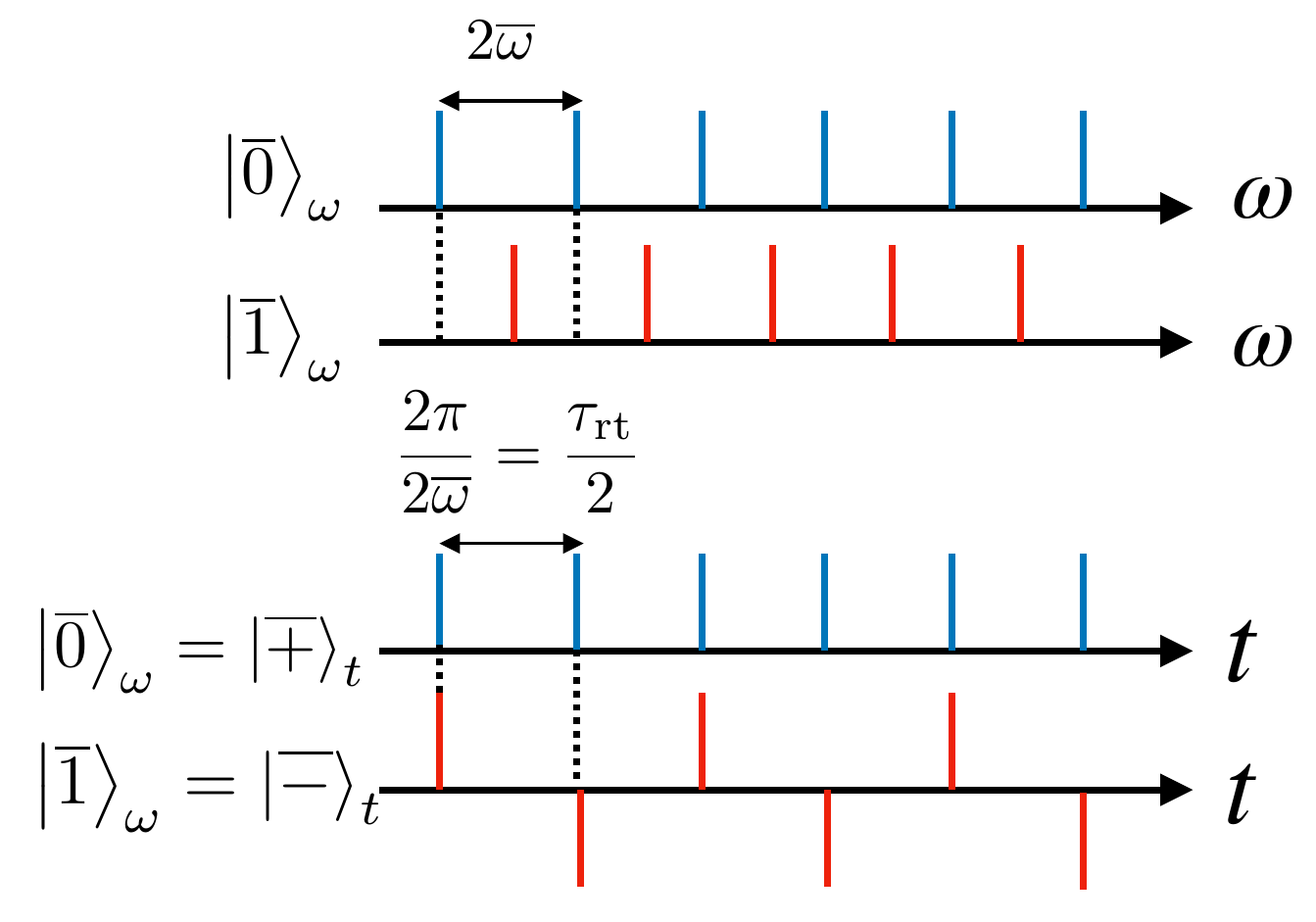}
\caption{\label{la11}Frequency-time GKP state in the frequency and time basis.}
\end{figure}

We start by providing the general framework to define single photon GKP states with many frequency modes. The two possible states of the qubit are the eigenstates of the displacement operators $\hat D(2\overline{\omega})$, ${\cal{\hat{D}}} (\frac{2\pi}{2\overline{\omega}})$, also called the stabilizer of the code (see Fig.\ref{la11}): 

\begin{equation}
\ket{\overline{0}}_{\omega}=\sum_{n\in\mathbb{Z}}\ket{\frac{\omega_{p}}{2}+2n\overline{\omega}},
\end{equation}
\begin{equation}
\ket{\overline{1}}_{\omega}=\sum_{n\in\mathbb{Z}}\ket{\frac{\omega_{p}}{2}+(2n+1)\overline{\omega}}.
\end{equation}
In the above, $2\overline{\omega}$ is the periodicity of the state and $\omega_{p}/2$ is a constant that will be later associated to a physical parameter. These states are called frequency-time square GKP states, because their frequency-time phase space representation is squared \cite{Performance}. For simplicity, we will call them simply time-frequency GKP states. 

Alternatively, as we show in Fig. \ref{la11},  we can use the time representation of the GKP states (up to normalization):
\begin{equation}
\ket{\overline{0}}_{\omega}=\tau_{\text{rt}}\sum_{n\in\mathbb{Z}}e^{i\frac{\omega_{p}}{2}\frac{n \tau_{\text{rt}}}{2}}\ket{\frac{n \tau_{\text{rt}}}{2}}=\ket{\overline{+}}_{t},
\end{equation}
\begin{equation}
\ket{\overline{1}}_{\omega}=\tau_{\text{rt}}\sum_{n\in\mathbb{Z}}e^{i\frac{\omega_{p}}{2}\frac{n \tau_{\text{rt}}}{2}}(-1)^{n}\ket{\frac{n \tau_{\text{rt}}}{2}}=\ket{\overline{-}}_{t},
\end{equation}
where $\tau_{\text{rt}}=\frac{2\pi}{\overline{\omega}}$. Here we have used that $\sum_{n\in\mathbb{Z}}e^{2in\pi t/\tau_{\text{rt}}}=\tau_{\text{rt}}\sum_{n\in\mathbb{Z}}\delta(t-n\tau_{\text{rt}})$. If we have the equality $\text{exp}(i\frac{\omega_{p}}{2}\frac{n \tau_{\text{rt}}}{2})=1$, which corresponds to the condition that  $\omega_{p}$ is a multiple of $\overline{\omega}$, we can define a qubit. It will be the case in the experimental configuration specified in the next section. For the sake of simplicity, we will now set $\omega_{p}=0$.
The $\ket{\overline{0}}_{t},\ket{\overline{1}}_{t}$ logical time GKP state (the eigenstates of the stabilizer $\hat{\cal{D}}(\tau_{\text{rt}})$  are then (up to normalization):
\begin{multline}
\ket{\overline{0}}_{t}=\tau_{\text{rt}}\sum_{n\in\mathbb{Z}}\ket{\frac{2n \tau_{\text{rt}}}{2}}=\ket{\overline{+}}_{\omega},\\
\ket{\overline{1}}_{t}=\tau_{\text{rt}}\sum_{n\in\mathbb{Z}}\ket{\frac{(2n+1) \tau_{\text{rt}}}{2}}=\ket{\overline{-}}_{\omega},\\
\end{multline}
where we have introduced $\ket{\overline{\pm}}_{t}=\frac{1}{\sqrt{2}}(\ket{\overline{0}}_{t}\pm\ket{\overline{1}}_{t})$ and analogously for $\ket{\overline{+}}_{\omega}$.

The frequency-time phase space representation of the frequency-time GKP state is analogous to the GKP state in the $(x,p)$ phase plane \cite{GKP}. The wave function of the coherent superposition  $\ket{\overline{+}}_{\omega}$, with the amplitude spectrum $S(\omega)=\sum_{n\in\mathbb{Z}}\bra{\omega}\ket{n\overline{\omega}}=\sum_{n\in\mathbb{Z}}\delta(\omega-n\overline{\omega})$ has the following chronocyclic Wigner distribution:
\begin{multline}
W(\omega,t)=\int_{\mathbb{R}} \text{d}\omega'e^{2i\omega' t}S(\omega-\omega')S^{*}(\omega+\omega')\\
= \sum_{n,m\in\mathds{Z}^{2}} (-1)^{nm}\delta(t-\frac{\pi}{\overline{\omega}}n)\delta(\omega-\frac{\omega_{p}}{2}-\frac{\overline{\omega}}{2}m).
\end{multline}

This shows that the chronocyclic Wigner distribution is negative when $n,m$ are both odd. Also, such states  are not physical since we are summing over an infinite number of perfectly well defined frequency (or time) modes.

\subsubsection{Physical frequency-time GKP state}\label{RealGKP}

In this section, we will see how to formally describe physical (intrinsically noisy) GKP states and how to physically interpret their number of peaks and the uncertainty of each mode. For that, we apply the formalism introduced in \cite{Knill}. 

Physical GKP states are constructed by applying a Kraus operator $\hat{\xi}$ to the ideal GKP state:
\begin{equation}\label{realGKP}
\ket{\tilde{0}}_{\omega} = \hat{\xi} \ket{\overline{0}}_{\omega} = \iint \text{d}\omega \text{d}t \xi(\omega,t) \hat{{\cal{D}}}(t)\hat{D}(\omega)\ket{\overline{0}}_{\omega}.
\end{equation}

If we suppose that frequency and time noise are uncorrelated Gaussian distribution, we have that: $\xi(\omega,t)=G_{\delta\omega}(\omega)G_{\kappa}(t)=e^{-\omega^{2}/2\delta\omega^{2}}e^{-t^{2}/2\kappa^{2}}$. The physical interpretation of these two Gaussian noises becomes clearer after performing the time integral in Eq. (\ref{realGKP}), which leads to: 
\begin{equation}\label{noisy1}
\ket{\tilde{0}}_{\omega}= \sum_{n\in\mathbb{Z}} \int T_{2n}(\omega) e^{-\omega^{2}\kappa^{2}/2} \ket{\omega}\text{d}\omega,
\end{equation}
with $T_{n}(\omega)=\text{exp}(-(\omega-n\overline{\omega})^{2}/(2\delta\omega^{2}))$. Time noise creates an envelope, limiting the number of relevant frequency modes while frequency noise introduces an intrinsic width to each peak.
Alternatively, we can construct the physical GKP state by permuting the time and frequency displacement operators. Since they are non-commuting operators, the state obtained by this procedure is not the same as Eq. (\ref{noisy1}). 

The physical origin of the finite width of the time and frequency distribution are errors that can be due to the propagation of single photons in a dispersive medium, as an optical fiber. They can also be related to time and frequency uncertainties inherent to the state preparation process, as is the case of the set-up  presented in Sec. \ref{sectionfour}.

\subsection{2D entangled time-frequency GKP state: some useful mathematical tools }

We now move to the two-qubit case and show how two GKP states can be entangled in time and frequency degrees of freedom. We start by considering an ideal separable two photon state, a two dimensional GKP state, that can be written as:
{{\small{\begin{multline}\label{Perfectgrid}
\ket{\overline{+}}_{\omega_{s}}\ket{\overline{+}}_{\omega_{i}}=\frac{1}{2}(\ket{\overline{0}}_{\omega_{s}}\ket{\overline{0}}_{\omega_{i}}+\ket{\overline{0}}_{\omega_{s}}\ket{\overline{1}}_{\omega_{i}}+\ket{\overline{1}}_{\omega_{s}}\ket{\overline{0}}_{\omega_{i}}\\+\ket{\overline{1}}_{\omega_{s}}\ket{\overline{1}}_{\omega_{i}}),
\end{multline}}}
As in the one qubit case, physical qubits can be constructed from (\ref{Perfectgrid}) using the noise model introduced previously, based on the application of  Kraus operators:

\begin{multline}\label{writtenbefore}
\ket{\tilde{+}}_{\omega_{s}}\ket{\tilde{+}}_{\omega_{i}}= \iint \text{d}t \text{d}t' \hat{{\cal{D}}}_{\text{s}}(t)\hat{{\cal{D}}}_{\text{i}}(t') G_{1/\Delta\omega_{p}}(t) G_{1/\Delta\omega_{-}}(t') \\
\cross \iint \hat{D}_{\text{s}}(\omega)\hat{D}_{\text{i}}(\omega') G_{\delta\omega_s}(\omega)G_{\delta\omega_i}(\omega')\text{d}\omega\text{d}\omega'\ket{\overline{+}}_{\omega_{s}}\ket{\overline{+}}_{\omega_{i}}.
\end{multline}
The functions $G_{\alpha}(x)$ are gaussians of width $\alpha$ corresponding to  the noise distribution on variables $x$. Frequency (or time) entanglement can be created by the application of a symmetric CNOT operator $\hat{C'}$  that performs the following operation:
\begin{equation}\label{CNOTGate}
\hat{C}'\ket{t_{s},t_{i}}=\ket{t_{s}+t_{i}}\ket{t_{s}-t_{i}}.
\end{equation}
Equivalently, $\hat{C}'\ket{\omega,\omega'}=\ket{\frac{\omega+\omega'}{2},\frac{\omega-\omega'}{2}}$. This gate operates in an analogous way to a balanced beam-splitter that acts on the frequency or time degree of freedom instead of the field's quadratures. 

The entangling gate $\hat{C}' $ tranforms displacement operators as follows: 
\begin{equation}\label{CNOTdispl}
\hat{C}'\hat{{\cal{D}}}_{\text{s}}(t)\hat{{\cal{D}}}_{\text{i}}(t')\hat{C'}^{-1}=\hat{{\cal{D}}}_{\text{s}}\left (\frac{t+t'}{2}\right )\hat{{\cal{D}}}_{\text{i}}\left (\frac{t-t'}{2}\right ).
\end{equation}

Using these definitions, we can now interpret the state produced from a by a nonlinear cavity producing photon pairs via SPDC in terms of GKP states and study its application to measurement based error correction.


\section{Production and manipulation of time-frequency grid state}\label{sectionfour}

We are now ready to apply the previous definitions to describe the spectrum of a photon pair generated by SPDC in a nonlinear optical cavity. We will focus on a specific platform in order to perform both a detailed numerical study and the experimental illustration of our results.

\subsection{Presentation of the integrated device for the generation of time-frequency GKP state}\label{fourA}

The device we consider here consists of an AlGaAs Bragg reflector waveguide emitting pairs of orthogonally polarised photons in the telecom band by type II SPDC. The working principle of the device is sketched in Fig.\ref{SPDCFIG} (a) \cite{amanti}. The device is pumped with a continuous wave laser at $\lambda_{p}=765$ nm having a linewidth $\Delta \omega_{p} = 2\pi \times100$ kHz, much smaller than the phase matching bandwidth and the free spectral range. This leads to the generation of strongly anticorrelated photon pairs over a spectral band of $2\pi \times 10.9$ THz centered around the frequency degeneracy as shown in the numerical simulations reported in Fig.\ref{SPDCFIG} (b). Moreover, the refractive index contrast between the semiconductor non-linear medium and the air induces a Fabry-Perot effect resulting in a built-in cavity.
The free spectral range of the cavity is $\overline{\omega}= 2\pi \times 19.2$ GHz, yielding  a comb like spectrum with approximatively 570 peaks. Fig.\ref{SPDCFIG} (c) shows the measurement of a portion of the Joint Spectral Intensity via Stimulated Emission Tomography \cite{Stimulated} evidencing a frequency comb structure.
  \begin{figure}[h!]
 \begin{center}
 \includegraphics[width=0.5\textwidth]{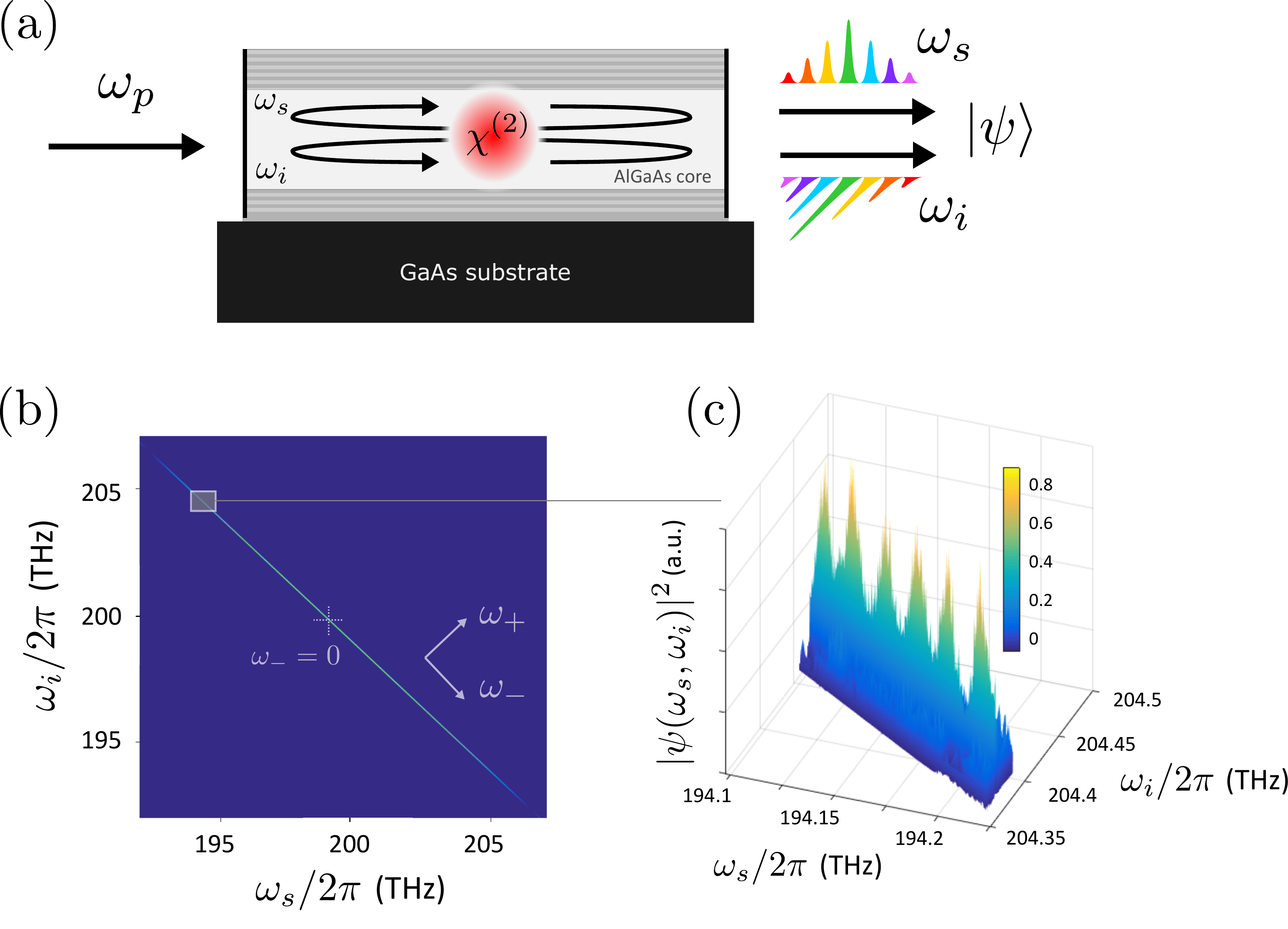}
 \caption{ \label{SPDCFIG} (a) A pump beam illuminates an AlGaAs waveguide where photon pairs are generated by SPDC. The refractive index contrast between AlGaAs and air creates a cavity around the nonlinear medium, as the waveguide's facets play the role of mirrors. (b) Simulated JSI of the state emitted by the nonlinear cavity, using the nominal structure of the device. (c) Experimental JSI (detail). } 
 \end{center}
\end{figure}

The generated two-photon state can be written as :
\begin{equation}\label{ProdState}
\ket{\psi}=\iint \text{d}\omega_{s}\text{d}\omega_{i} \text{JSA}(\omega_{s},\omega_{i})\ket{\omega_{s}}\ket{\omega_{i}},
\end{equation}

where the JSA is the product of four terms:
\begin{equation}\label{JSAspdc}
\text{JSA}(\omega_{s},\omega_{i})=f_{+}(\omega_{+})f_{-}(\omega_{-})f_{\text{cav}}(\omega_{s})f_{\text{cav}}(\omega_{i}).
\end{equation}

We have defind $\omega_\pm=\frac{\omega_s\pm\omega_i}{2}$.The  function $f_{+}$ is related to the conservation of the energy, while $f_{-}$ is related to the phase matching condition. Both functions can be modeled as Gaussian functions. The effect of  the cavity on each mode (signal and idler), is taken into account by the cavity functions $f_{\text{cav}}(\omega_{s (i)})$ that act as an imperfect frequency filter.
We will assume that the cavity function is a sum of Gaussians, which is a good approximation in the limit of a high finesse cavity:
\begin{equation}
f_{\text{cav}}(\omega)=\sum_{n\in\mathbb{Z}}T_{n}(\omega),
\end{equation}
where again $T_{n}(\omega)=\text{exp}(-(\omega-n\overline{\omega})^{2}/(2\delta\omega^{2}))$.
The frequency width $\delta\omega$ depends on the finesse of the cavity. For a high finesse cavity, we have $\overline{\omega} \gg \delta\omega$, and the two photon state can be written as:
{\small
\begin{equation}\label{SPDC}
\ket{\psi}=\sum_{n,m} \iint \text{d}\omega_{s}\text{d}\omega_{i} f_{+}(\omega_{+})f_{-}(\omega_{-})T_{n}(\omega_{s})T_{m}(\omega_{i})\ket{\omega_{s}}\ket{\omega_{i}}.
\end{equation}}

The resulting state is analogous to a grid state because of the cavity functions. We can identify

\begin{eqnarray}\label{Ts}
&&\sum_{n,m} \iint \text{d}\omega_{s}\text{d}\omega_{i} T_{n}(\omega_{s})T_{m}(\omega_{i})\ket{\omega_{s}}\ket{\omega_{i}}= \\
&& \iint \hat{D}_{\text{s}}(\omega)\hat{D}_{\text{i}}(\omega') G_{\delta\omega}(\omega)G_{\delta\omega}(\omega')\text{d}\omega\text{d}\omega'\ket{\overline{+}}_{\omega_{s}}\ket{\overline{+}}_{\omega_{i}}, \nonumber
\end{eqnarray}
which, as mentioned before, represents GKP states affected by frequency noise. Completing the description of state (\ref{SPDC}) requires considering also time displacement operators representing noise in the time variables as follows: 

{\small{\begin{multline}\label{WithD}
\ket{\psi}=\iint \hat{{\cal{D}}}_{\text{s}}\left (\frac{t+t'}{2}\right )\hat{{\cal{D}}}_{\text{i}}\left (\frac{t-t'}{2}\right )G_{1/\Delta\omega_{p}}(t)G_{1/\Delta\omega_{-}}(t')\text{d}t\text{d}t'\\
\times \iint \hat{D}_{\text{s}}(\omega)\hat{D}_{\text{i}}(\omega') G_{\delta\omega}(\omega)G_{\delta\omega}(\omega')\text{d}\omega\text{d}\omega'\ket{\overline{+}}_{\omega_{s}}\ket{\overline{+}}_{\omega_{i}}.
\end{multline}}}
In this expression, we considered the possibility of having different widths in the time noise distribution, given by $\Delta \omega_-$ and $\Delta \omega_p$. Indeed such widths are related to the phase matching condition and to energy conservation, respectively, creating envelopes in the axis $\omega_-$ and $\omega_+$,  limiting the dimension of the produced grid.

\begin{figure}[h]
\begin{center}\label{JS1}
 \includegraphics[scale=0.48]{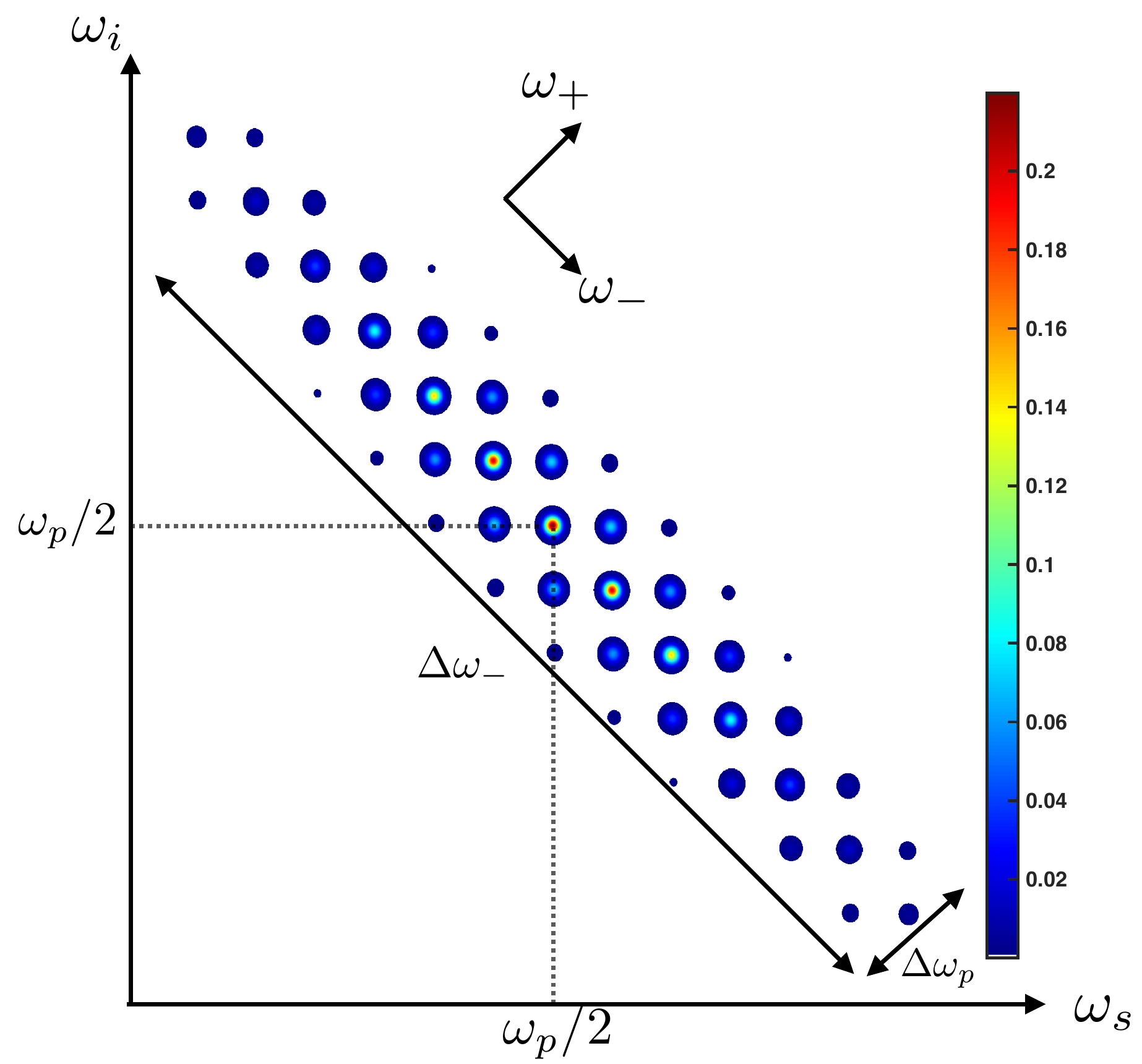}
 \caption{\label{JSIlarge} Numerical simulation of the Joint Spectral Intensity for a two photon source in an optical cavity with arbitrary units. The size of the ellipse is delimited by the energy conservation (with a frequency width $1/\Delta\omega_{p}$) and the phase matching condition (with a frequency width of $1/\Delta\omega_{-}$). The center of the ellipse corresponds to the frequency degeneracy. Here the state is said to be anti-correlated since  $\Delta\omega_{-} > \Delta\omega_{p} $.}
 \end{center}
\end{figure}

Applying the displacement operators in (\ref{WithD}) leads to:
{\small
\begin{multline}
\ket{\psi}=\iint \text{d}t\text{d}t' \text{d}\omega\text{d}\omega' G_{1/\Delta\omega_{p}}(t)G_{1/\Delta\omega_{-}}(t')G_{\delta\omega}(\omega)G_{\delta\omega}(\omega')\\
\times \ \sum_{n,m}e^{i(n\overline{\omega}+\omega)\frac{(t+t')}{2}}e^{i(m\overline{\omega}+\omega')\frac{(t-t')}{2}}\ket{n\overline{\omega}+\omega}\ket{m\overline{\omega}+\omega'},
\end{multline}}
and, by integrating over time and performing a change of variable, we obtain:

{\small{
\begin{multline}\label{SPDC1}
\ket{\psi}=\sum_{n,m\in\mathds{Z}^{2}} \iint \text{d}\omega_{s}\text{d}\omega_{i} G_{\Delta\omega_{p}}(\omega_{+})G_{\Delta\omega_{-}}(\omega_{-})\\
\cross T_{n}(\omega_{s})T_{m}(\omega_{i})\ket{\omega_{s}}\ket{\omega_{i}},
\end{multline}}}
which has exactly the same form as Eq. (\ref{ProdState}) and Eq. (\ref{SPDC}), justifying the interpretation in terms of GKP states as depicted in  Fig. \ref{JSIlarge}. On Fig. \ref{CNOT1}, we recap in a quantum circuit representation all the time-frequency gates which act on the initial ideal GKP state.

The states described by Eq. (\ref{SPDC1})  are a noisy frequency entangled states, displaying an elliptical JSI in the in the $(\omega_{+},\omega_{-})$ basis. The ellipticity $R$ of the JSI determines how entangled the state is. It is defined as $R=\frac{1/\Delta\omega_{-}^{2}-1/\Delta\omega_{p}^{2}}{1/\Delta\omega_{-}^{2}+1/\Delta\omega_{p}^{2}}$, and the time noise in axis $\omega_-$ and $\omega_+$  plays a role in the correlation/anti-correlation of the photons.  A state with an arbitrary positive ellipticity is represented on the numerical simulation Fig. \ref{JSIlarge}. Note that such biphoton state can be produced with an integrated chip in a transverse pump configuration \cite{Stimulated,tapis}. Thus, we can define $\ket{\psi}= \widetilde {\ket{\bar{+}}\ket{\bar {+}}}$, since state $\ket{\psi}$ can be interpreted as two GKP states entangled by noise.

It is interesting to notice that in the case where $G_{\Delta\omega_{p}}(\omega_{+})=\delta(\omega_{+})$ and  $\overline{\omega}/\delta\omega \ll 1$, we can approximate $\ket{\psi}= \widetilde {\ket{\bar{+}}\ket{\bar {+}}} \simeq \hat C' \ket{{\tilde { +}}}\ket{{\tilde { +}}}$ where we used Eq. (\ref{writtenbefore}) and  (\ref{CNOTGate}). In this case, $\ket{\psi}$ is anti-correlated,  and $\Delta\omega_{p}\ll \Delta\omega_{-}$. The JSI is shown in Fig. \ref{SPDCFIG}(b) and it is close to  a line along the $\omega_{-}$ direction.  We thus have that  $\text{JSA}(\omega_{s},\omega_{i})\simeq\delta(\omega_{+}-\omega_{p})f_{-}(\omega_{-})f_{\text{cav}}(\omega_{s})f_{\text{cav}}(\omega_{i})$ and  integration over $\omega_{+}$ in Eq. (\ref{SPDC}) leads to :
\begin{multline}\label{SPDCc}
\ket{\psi}=\int \text{d}\omega_{-} f_{-}(\omega_{-})f_{\text{cav}}\left (\frac{\omega_{p}+\omega_{-}}{2}\right )f_{\text{cav}}\left (\frac{\omega_{p}-\omega_{-}}{2}\right )\\ \cross \ket{\frac{\omega_{p}+\omega_{-}}{2},\frac{\omega_{p}-\omega_{-}}{2}}.
\end{multline}
and this state is the one produced experimentally whose JSI is represented on Fig. \ref{SPDCFIG}.

\begin{figure}[h]
\begin{center}\label{CNOTFigure}
 \includegraphics[scale=0.6]{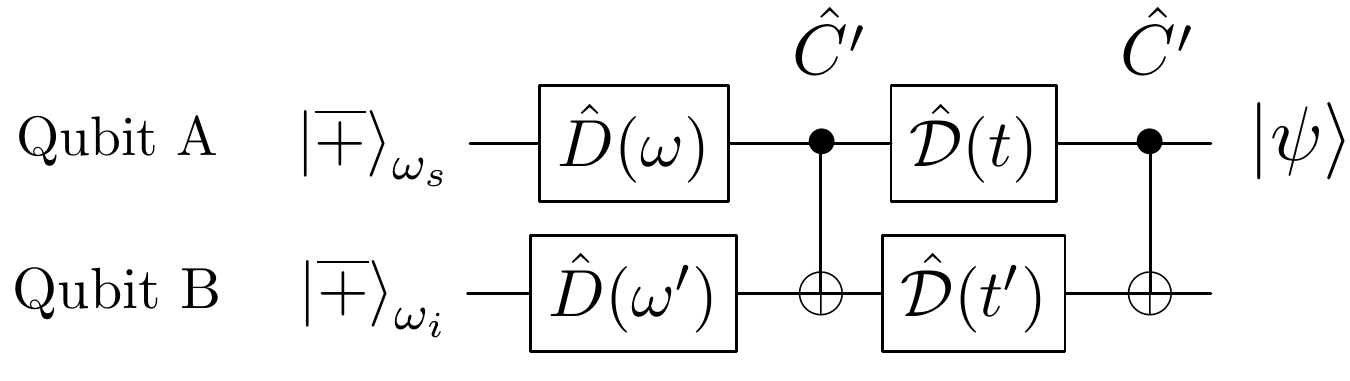}
 \caption{\label{CNOT1} Quantum circuit representing the generation of entangled 2D time-frequency GKP state. }
 \end{center}
\end{figure}

\subsection{Applications}\label{fourB}

The previously described  2D entangled time-frequency GKP state can be used  to implement a measurement-based error correction protocol, which was previously defined for quadrature position-momentum continuous variable \cite{Menicuccierror}. 

In this scenario, the result of a measurement performed on one qubit (say, B, encoded in the idler photon, also called the ancilla) is used to correct the error on the other qubit (A, encoded in the signal photon, also called the data qubit).  

We will consider the effect of a time measurement on the ancilla qubit of state Eq. (\ref{SPDC1}). Since both qubits are entangled, measuring the ancilla qubit (B) has an effect on the data qubit (A), as shown in Fig. \ref{MBQC}  \cite{Steane}. The operation realized in qubit B is teleported to qubit A, up to a known displacement on qubit A, which is given by the result of the measurement performed in qubit B. In the spirit of QEC, the interest of this approach is that, if noise corresponds to displacements in conjugate variables, as it is the case in the GKP code, one can show that,  if qubit B is measured in one variable (time or frequency), its error in the measured variable is teleported to qubit A's error in the same variable. Thus, if qubit B's error is smaller than A's, this scheme can be used to decrease the noise in physical GKP states of A \cite{Menicuccierror, Tomthesis}.

\begin{figure}[h]
\begin{center}
 \includegraphics[scale=0.65]{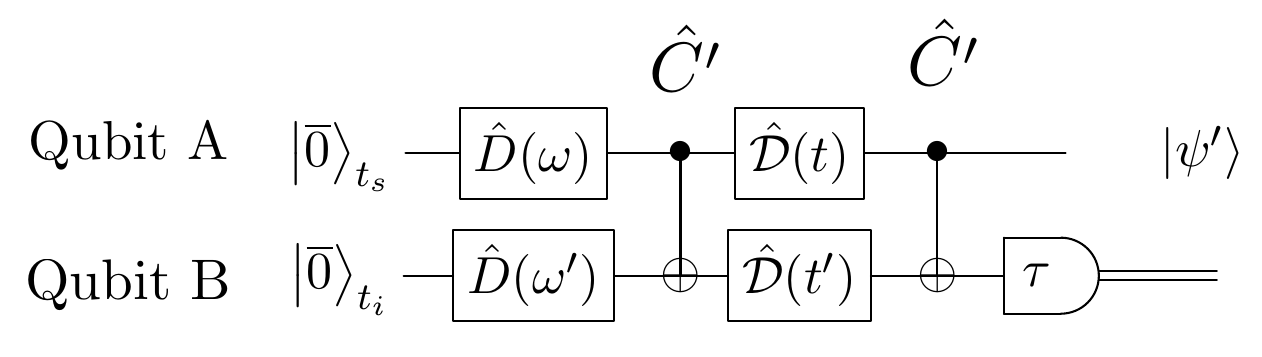}
 \caption{ \label{MBQC}The data qubit (signal) in arm A and the ancilla one  (idler) in arm B are prepared in state $\ket{\bar +}_{\omega_{s} }\ket{\bar +}_{\omega_{i}}=\ket{\overline{0}}_{t_{s}}\ket{\overline{0}}_{t_{i}}$. After displacements and the $\hat C'$ gate, we perform a time measurement on the ancilla.}
 \end{center}
\end{figure}

An interesting aspect of using measurement-based techniques is that they provide an alternative to a deterministic two-qubit gate in single photon-based devices. As a matter of fact, implementing deterministic gates is a challenge in such set-ups, and starting from useful entangled resources can help achieving determinacy in different protocols. Possible ways to scale up the generation of time-frequency GKP states would be using on-demand production of pure single photon states, reviewed in \cite{Senellart} for instance. Such ideas can be combined to implement efficient frequency gates, which are currently realised with electro-optic modulators and pulse-shapers  \cite{Lukens, Lu1}. 

We expect that the fast technological evolution of the integrated circuits physics will enable effective photon-photon interaction with higher probability in the near future.

\subsection{Experimental manipulation of time-frequency grid state and state detection}\label{fourC}

 In this section, we study the implementation of a single qubit gate ${\bf \hat{Z}}_{t_s}$ on state  Eq. (\ref{SPDCc}).

A possibility to manipulate frequency states is using electro-optical modulation (EOM) \cite{Lu}  for frequency-bin encoded qubits. Such techniques can also be used in the present context, with the difference that while in Ref.\cite{Lu} each frequency is manipulated independently, in the present encoding redundancy is a key aspect, and qubit manipulation requires acting on the whole frequency comb. It must then be manipulated as a whole, a situation that does not add any experimental complexity to the techniques considered in \cite{Lu}. 
 \begin{figure}[h!]
 \includegraphics[width=0.4\textwidth]{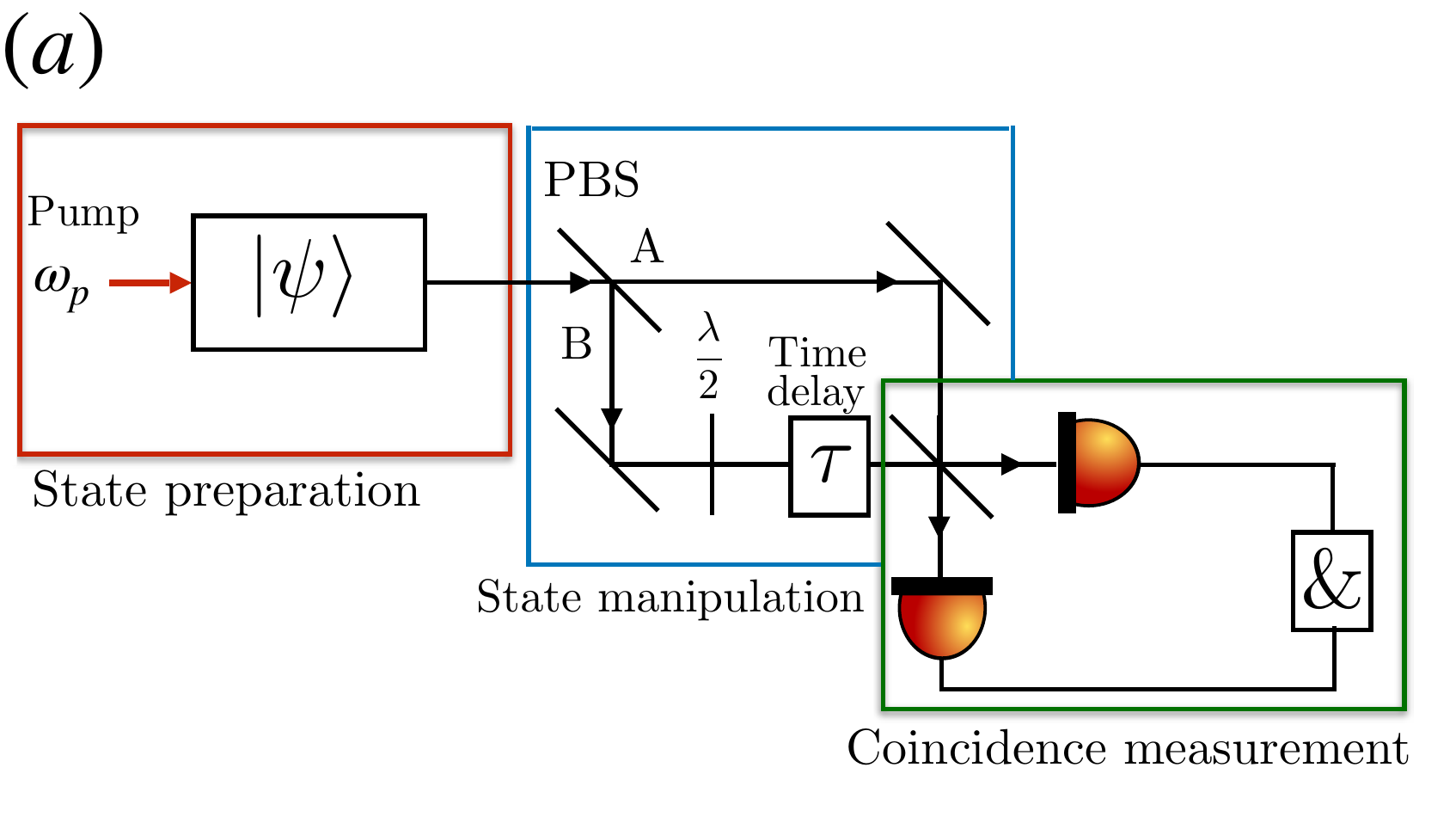}
  \includegraphics[width=0.45\textwidth]{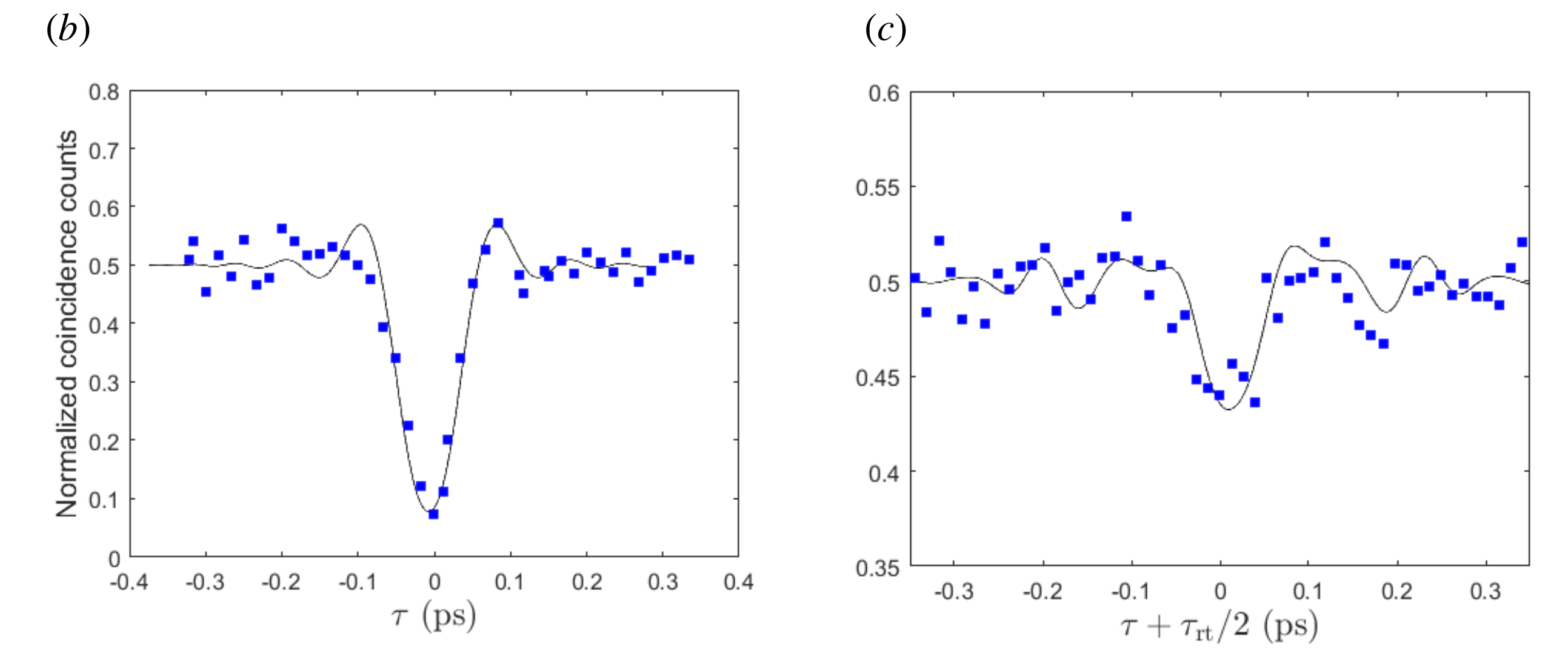}
 \caption{ \label{HOM} (a) Hong-Ou-Mandel experiment enabling state manipulation and measurement.  After being generated, signal and idler photons are separated and sent into two different arms of a HOM interferometer with a polarizing beam-splitter (PBS). Time delay ($\tau$) in one arm performs a ${\bf \hat{Z}}_{t_{s}}$ gate for $\tau=-\tau_{\text{rt}}/2$. In order to have the same polarization for the photons, a half-wave plate is added. State measurement can be done by recombining both photons in a second beam splitter and performing coincidence measurements for different values of $\tau$. (b) Experimental coincidence measurements corresponding to the state $\widetilde {\ket{\bar{+}}\ket{\bar {+}}}$. (c) Experimental coincidence measurements corresponding to the state $\widetilde {\ket{\bar{-}}\ket{\bar {+}}}$. The continuous lines are  the result of numerical calculations taking into account the effects of birefringence, reflectivity and chromatic dispersion in the AlGaAs chip.} 
\end{figure}

Interestingly, using EOM is not strictly necessary to manipulate time-frequency GKP states. We demonstrate here an experimentally simpler way to implement a quantum gate ${\bf \hat{Z}}_{t_s}$ for time-frequency GKP states  and obtain a signature of the manipulation using a Hong-Ou-Mandel  (HOM) interferometer \cite{HOM,LO}, that can be used for state measurement, as detailed in the following. The HOM setup is sketched in Fig. \ref{HOM} (a): signal and idler photons are sent to different arms of an interferometer, A and B. We consider to be in the limit  $\Delta\omega_{p}\ll \Delta\omega_{-}$ for simplicity.
Introducing a time delay $\tau$ between the two arms, the two photons acquire a phase difference such that the biphoton state arriving in the recombining beam-splitter is given by:
\begin{multline}\label{SPDCtime}
\ket{\psi(\tau)}=\int \text{d}\omega_{-} e^{-i(\omega_{-}+\omega_{p})\tau/2}f_{-}(\omega_{-})f_{\text{cav}}\left (\frac{\omega_{p}+\omega_{-}}{2}\right )\\ \cross f_{\text{cav}}\left (\frac{\omega_{p}-\omega_{-}}{2}\right )\ket{\frac{\omega_{p}+\omega_{-}}{2},\frac{\omega_{p}-\omega_{-}}{2}}.
\end{multline}
Without loss of generality for the present purposes, we consider $g(\omega_-)=f_{-}(\omega_-)f_{\rm cav}( \frac{\omega_{p}+\omega_{-}}{2})f_{\rm cav}( \frac{\omega_{p}-\omega_{-}}{2} )$ to be real. This function is also symmetric with respect to $\omega_-=0$.  The phase $e^{-i\omega_-\tau}$ corresponds to a displacement of $\tau$ in time, the conjugate variable to $\omega_-$, as shown in Sec. \ref{sectiontwo}. It corresponds to the application of the $\hat{\cal{D}}_{s}(\tau)$ operator to the signal photon {\it before} the entangling operation.

 By choosing  $\tau=-\tau_{\text{rt}}/2$  the $n$-th peaks of  $g(\omega_-)$ with $n$ even, remain unchanged, while for $n$ odd, they gain a $\pi$ phase and change signs, implementing the gate ${\bf \hat{Z}}_{t_s}\ket{ \bar{ +}}_{\omega_{s}}=\ket{\bar { -}}_{\omega_{s}}$ with a simple interferometric configuration and coincidence detection. Consequently, before its arrival on the HOM recombining beam-splitter, the two-photon state can be written as $\ket{\psi'}= \widetilde {\ket{\bar{-}}\ket{\bar {+}}}$.


The signature of  time displacement operator  and the orthogonality of the two states can be detected by measuring temporal correlations with a HOM interferometer. As shown in \cite{Tom, Guillaume}, the HOM experiment is a direct measurement of the chronocyclic Wigner distribution of the phase-matching part of the JSA. The first experimental demonstration of these ideas can be found in \cite{WignerExp}. In the experimental context discussed here, it gives access to a cut in the time-frequency phase space of the Wigner function associated to the global variable  $\omega_-$, $W(\mu, \tau)$, where $\mu$ is the amplitude of displacement of  $\omega_-$  and $\tau$ the amplitude of displacement in time. The HOM experiment corresponds thus to a cut along the $\mu=0$ line, where $\tau$ is varied. The partial information obtained is enough to distinguish between the two orthogonal states.

We have implemented the setup of Fig.\ \ref{HOM}\ (a) on the state produced by our AlGaAs device presented  in \ref{fourA}. For $\tau = 0$,   we expect a coincidence dip with a visibility fixed by the degree of indistinguishability between the signal and idler photons: this corresponds to the state $ \hat C'\ket{\tilde { +}}_{\omega_{s}}\ket{\tilde { +}}_{\omega_{i}}$ . For $\tau = -\tau_{\text{rt}}/2$, we expect to observe a replica of the previous dip whose visibility is related to a combination of facets reflectivity, birefringence and chromatic dispersion: this corresponds  to the state $\hat C' {\bf \hat{Z}}_{t_s}\ket{\tilde { +}}_{\omega_{s}}\ket{\tilde { +}}_{\omega_{i}}$. The results of the corresponding measurements are shown respectively in Fig.\ \ref{HOM}\ (b) and (c). In the first (Fig.\ \ref{HOM}\ (b),  in the vicinity $\tau=0$) the visibility is $86\%$, while in the second case (Fig.\ \ref{HOM}\ (c), around $\tau=-\tau_{\text{rt}}/2$) we obtain a visibility of 12\%, making these two states well distinguishable. 
We perform  in Fig.~\ref{visibility} numerical simulations of the visibility of the second dip of the HOM experiment as a function of the cavity reflectivity and for different bandwidths of the filters placed before the beam-splitter.
 The intersection of the dashed lines indicates the conditions in which the experiment has been performed: a modal reflectivity of the facets of 0.3 without frequency filters, which leads to a theoretical prediction of 15\% of visibility, which is in good agreement with the experimentally observed result of  12\%, see Fig.\ref{HOM} (c). Such visibility is enough to distinguish both possible GKP states.  Note that the visibility of the second dip can be increased by depositing a reflective coating on the facets, but this solution would equally enhance the detrimental effect of the cavity birefringence by making peaks corresponding to different polarizations more and more distinguishable.  In addition to coating, a potential strategy to improve the visibility is to add frequency filters in order to select the central part of the spectrum and reduce the effect of birefringence and chromatic dispersion. For instance, the total frequency bandwidth for about 500 peaks is 70 nm as shown in Fig.~\ref{visibility}. For each curve in Fig.~\ref{visibility}, we note that the visibility reaches a maximum and then decreases when increasing the reflectivity, well illustrating our discussion on the antagonist roles of the reflectivity and the birefringence. It shows that a visibility of the order of 80 \% is well on reach. Numerical and analytical results are presented in the Appendix \ref{HOMGATES} for such values of reflectivity.   \\

In summary, a visibility higher than 80\% for the adjacent dips from the central dip can be obtained by improving the reflectivity of the facets and filtering the produced spectrum, which would decrease the number of exploitable peaks and reduce the detection rate but would still keep them of the order of a hundred.\\

Further possible manipulation of the time-frequency grid state has been proposed in \cite{amanti}, where tuning the pump frequency allows to engineer the JSA symmetry. Compared to other schemes where each comb line is manipulated individually \cite{Lu1,Lu2},  this technique enables to address the odd and the even peaks individually with low optical losses. 
In the formalism developed in Sec. \ref{firstsection}, the pump tuning corresponds to a frequency displacement operation on the signal and idler photon, see Fig. 2 of Ref. \cite{amanti}, namely to non-linear operation without the need to introduce a non-linear material or EOM after the generation of the two-photon state.

\begin{figure}[h!]
\begin{center}
 \includegraphics[scale=0.35]{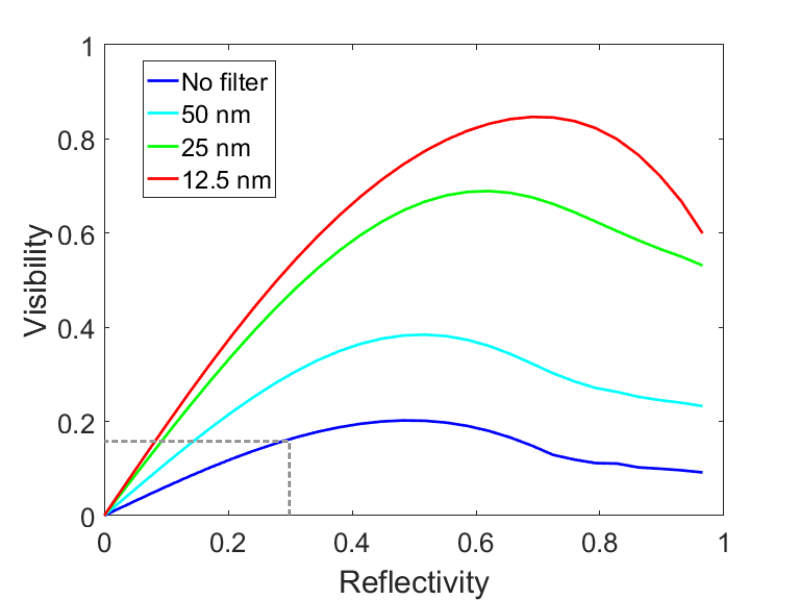}
 \caption{ \label{visibility}Numerical simulation of the visibility of the secondary peaks, nearest to the central dip,  as a function of the reflectivity of the facets for different bandwidth of the frequency filters placed before the beam-splitter. The birefringence and the chromatic dispersion are here taken into account. The intersection of the two dashed lines indicates the conditions of the realized experiment whose coincidence measurement is presented in Fig \ref{HOM}.(c).}
 \end{center}
\end{figure}

\subsection{Experimental proposal for quantum error correction in time-frequency variable}\label{sectionMBQC}

 We now discuss the experimental feasibility of the time quantum error correction. The Joint Temporal Intensity of the state given by Eq. (\ref{SPDC}) is represented in Fig. \ref{JTA}. The state is periodic, with periodicity of $2\pi/\overline{\omega}=50$ ps using the parameters of the above described source, along the two orthogonal directions $t_{\pm}$. But since the inverse of the energy conservation width $1/\Delta\omega_{+}$ is much larger than the cavity round trip time $\tau_{\text{rt}}$, the periodicity along the $t_{+}$ is not visible. A time measurement of the idler photon leads to a random temporal distribution which corresponds to the different peaks along the $t_{-}$ axis. A single photon detector should have $50$ ps temporal resolution to distinguish these peaks, which is possible with nowadays technology. It corresponds to what is commonly designated as photon heralding. Once the state is measured, further time or frequency correction operators could be applied on the data \cite{Knill}. \\

\begin{figure}[h!]
\begin{center}
 \includegraphics[scale=0.4]{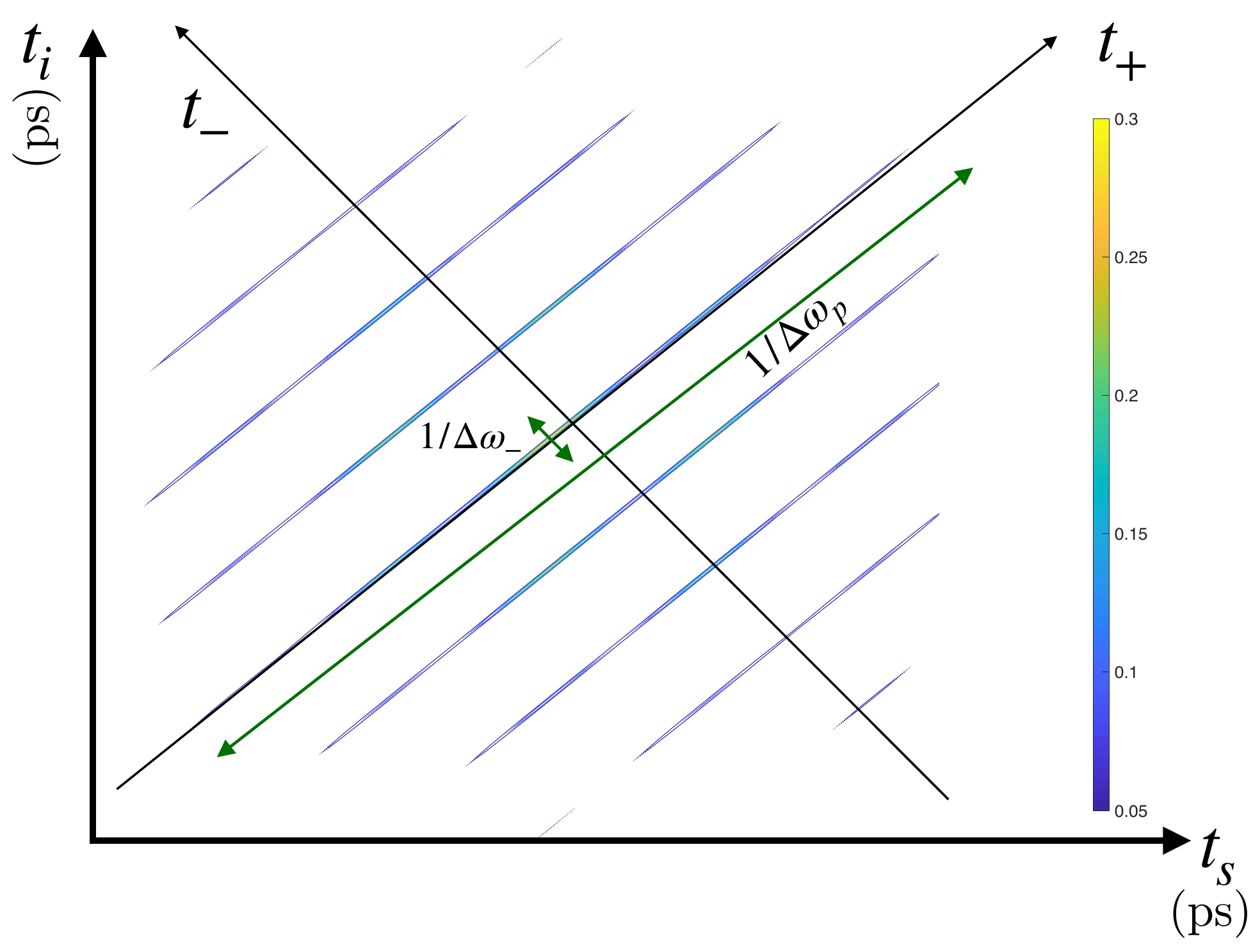}
 {\small{\caption{ \label{JTA}Numerical Simulation of the Joint Temporal Intensity of the time-frequency GKP state in the case $\Delta\omega_{-} \gg \Delta\omega_{p}$ with a 50 ps periodicity. It corresponds to the Fourier transform of the state shown in Fig. \ref{SPDCFIG}.(b). The state is periodic in both directions, but since the $1/\Delta\omega_{p}\gg 2\pi/\overline{\omega}$, we can not see the periodicity in the $t_{+}$ direction since the data qubit is very noisy. \\} }}
 \end{center}
\end{figure}

Error correction is also possible in frequency degrees of freedom, and it requires measuring one of the photons in the $\omega_{\pm}$ variables. This operation could be performed with non-linear devices implementing a controlled quantum gate in the frequency degrees of freedom. \\

\section{Conclusion}
As a conclusion, we detailed a formalism establishing the analogy between continuous variables consisting of many modes of single photons and those associated to a single mode field's quadratures.  We introduced a rigorous construction and provided a physical and mathematical meaning to the chronocyclic Wigner distribution in a quantum optics experiment. Using the introduced formal analogies, we showed that experimental setups consisting of a  SPDC source and a filtering cavity can be a natural source of time-frequency GKP states, which are time-frequency non-Gaussian states useful for fault-tolerant quantum protocols manipulating continuous variables. Qubits can be encoded in frequency and time degrees of freedom of photons and entangled GKP states can be generated and manipulated. We have experimentally illustrated these results in an integrated optical platform. Finally, we have shown that the produced state is a resource for MBQC and error correction, and both can be implemented through time or frequency measurement of one photon of the pair. A natural perspective is to combine our results with already existing technologies for quantum photonic circuits for further applications and scaling \cite{waveguideint2,kues,wangmulti,dietrich}.
Using the measurement based principle, other protocols involving quadrature position-momentum GKP state could be generalised to time-frequency variables such as the correction of Gaussian errors using GKP state as a non-Gaussian ressource with the two-mode GKP repetition code \cite{correction,correction2}. 

\section*{ACKNOWLEDGMENT}

The authors gratefully acknowledge ANR (Agence Nationale de la Recherche)  for  the  financial  support  of this work through Project SemiQuantRoom (Project No.ANR-14-CE26-0029) and through Labex SEAM (Science and  Engineering  for  Advanced  Materials  and  devices) project  ANR-10-LABX-0096 et ANR-18-IDEX-0001. The French  RENATECH  network  and  Universit\'e Sorbonne Paris Cit\'e  for PhD fellowship to G.M. are also warmly acknowledged. A. Ketterer acknowledges support by the Georg H. Endress foundation.

\appendix

\section{IMPLEMENTING GATES USING THE HONG-OU-MANDEL EXPERIMENT}\label{HOMGATES}

In this section, we detail how to implement the single qubit gate ${\bf \hat{Z}}$ for the frequency-time GKP state. The frequency $\omega_{p}$ is taken different from zero.
For simplicity, we will describe the principle of the gate for an ideal GKP state. Starting from the Eq. (\ref{SPDCc}) and supposing that each photon goes to one arm of a Hong-Ou-Mandel (HOM) interferometer where a linear medium was inserted in one of the arms (see Fig. \ref{HOM}), the wave function can be written, after the beam-splitter, taking into account only the coincidence terms as,
{\small
\begin{multline}
\ket{\psi}_{\tau}=\frac{1}{2}\int \text{d}\omega_{-} f_{-}(\omega_{-})f_{\text{cav}}(\frac{\omega_{p}+\omega_{-}}{2})f_{\text{cav}}(\frac{\omega_{p}-\omega_{-}}{2})\\\cross e^{-i\frac{(\omega_{p}+\omega_{-})\tau}{2}}(\ket{\frac{\omega_{p}+\omega_{-}}{2},\frac{\omega_{p}-\omega_{-}}{2}}-\ket{\frac{\omega_{p}-\omega_{-}}{2},\frac{\omega_{p}+\omega_{-}}{2}}),
\end{multline}}

where $\tau$ is the temporal delay introduced by the linear medium in the upper path. After performing a change of variable, we obtain:
{\small
\begin{multline}
\ket{\psi}_{\tau}=\frac{1}{2}\int \text{d}\omega_{-} (f_{-}(\omega_{-})e^{-i\frac{\omega_{-}\tau}{2}}-f_{-}(-\omega_{-})e^{+i\frac{\omega_{-}\tau}{2}})\\f_{\text{cav}}(\frac{\omega_{p}+\omega_{-}}{2})f_{\text{cav}}(\frac{\omega_{p}-\omega_{-}}{2})\ket{\frac{\omega_{p}+\omega_{-}}{2},\frac{\omega_{p}-\omega_{-}}{2}}.
\end{multline}}
In the last equation, we discard an unimportant global phase. The coincidence probability $I(\tau)=\iint \text{d}\omega_{s}\text{d}\omega_{i} \abs{\bra{\omega_{s},\omega_{i}}\ket{\psi}_{\tau}}^{2}$ reads:
\begin{multline}
I(\tau)=\frac{1}{2}[1-\frac{1}{N}\text{Re}(\int \abs{f_{\text{cav}}(\frac{\omega_{p}+\omega_{-}}{2})f_{\text{cav}}(\frac{\omega_{p}-\omega_{-}}{2})}^{2}\\\cross f_{-}(\omega_{-})f_{-}^{*}(-\omega_{-})e^{-i\omega_{-}\tau/2}\text{d}\omega_{-})],
\end{multline}
where $N=\int \abs{f_{\text{cav}}(\frac{\omega_{p}+\omega_{-}}{2})f_{\text{cav}}(\frac{\omega_{p}-\omega_{-}}{2})}^{2}f_{-}(\omega_{-})\cross \\ f_{-}^{*}(-\omega_{-})\text{d}\omega_{-}$. In \cite{Douce}, it was shown that the coincidence probability is proportional to a cut of the chronocyclic Wigner distribution at $\omega_{-}=0$. We can then only partially characterise the state, but the information obtained is enough to analyse different time-frequency GKP states.

Experimentally we realize the HOM experiment having as initial state (\ref{SPDCc}), so that we can observe the effect of the gate ${\bf \hat{Z}_{t_{s}}}$. For $\tau=-\tau_{\text{rt}}/2$, the odd peaks of state  (\ref{SPDCc}) gain a negative amplitude, which means that  a ${\bf \hat{Z}_{t_{s}}}$ gate was implemented. This corresponds to the transformation $ {\bf \hat{Z}}_{t_s}\ket{\bar { +}}_{\omega_{s}}$ applied to the ideal GKP state, which means that the physical state is transformed according to  $\widetilde {\ket{\bar{+}}\ket{\bar {+}}}\rightarrow  \widetilde {\ket{\bar{-}}\ket{\bar {+}}}$. 

Here, we perform the analytical calculation for the coincidence probability of the state, for a high reflectivity of the cavity and without taking into account the birefringence and the chromatic dispersion. Assuming that $\overline{\omega}/\delta\omega \gg1$, we have:
{\small
\begin{equation}
I(2\tau)=\frac{1}{2}[1-e^{-\tau^{2}\delta\omega^{2}/2}\sum_{n=-d}^{d}\alpha_{n}\text{cos}(n\overline{\omega}\tau)],
\end{equation}}
where $d$ is the number of peaks, $\frac{1}{N}=\frac{1}{\sum_{n=-d}^{d}\alpha_{n}}$ and $\alpha_{n}=e^{-(\frac{\omega_{p}}{2}-n\overline{\omega})^{2}/\delta\omega^{2}}$. In Fig. \ref{DIP} (a), (b), we show the plot of the coincidence probability with arbitrary units for a cavity with a reflectivity of $r=0.9$. The HOM interference exhibits replica \cite{LO}, and depending on the time displacement we perform, we obtain for the signal photon, the state $\ket{\tilde{+}}_{\omega_{s}}$or $\ket{\tilde{-}}_{\omega_{s}}$. The visibility of the central dip and the nearest replica are too close to distinguish the two states, contrary to the experimentally studied case where low reflectivity and chromatic dispersion increase the coincidence probability of the state $\ket{\tilde{-}}_{\omega_{s}}$. For a high reflectivity of the cavity, to distinguish the two orthogonal states, we then have to choose two replicas away from the central dip.

\begin{figure}[h!]
\begin{center}
 \includegraphics[scale=0.55]{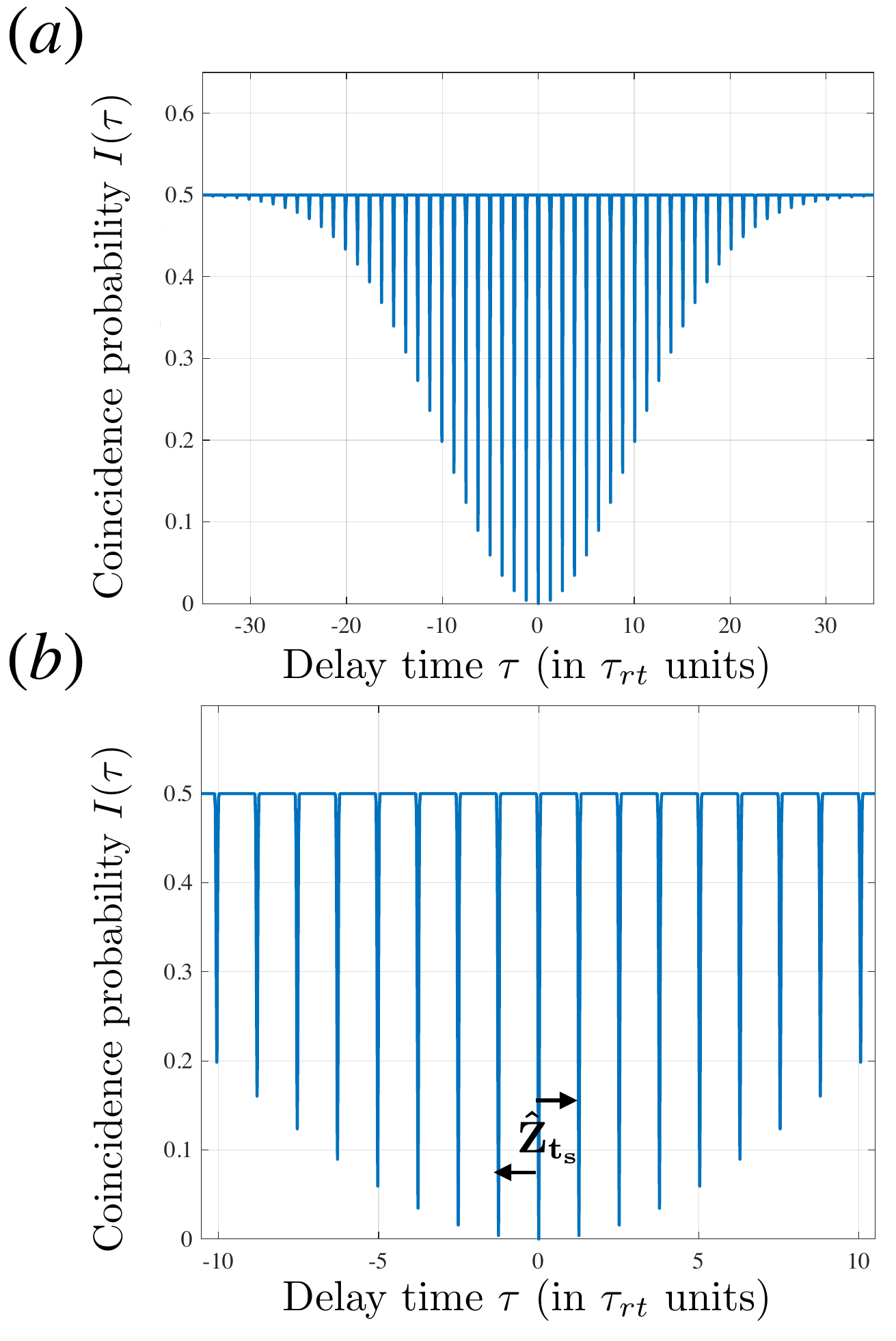}
 \caption{\label{DIP} (a) Numerical simulation of the HOM experiment for the two photon state for a highly reflective cavity without taking into account the birefringence and the chromatic dispersion. Coincidence probability as a function of the delay in units of $\tau_{\text{rt}}$. Selecting  $\tau=\pm\tau_{\text{rt}}/2$ performs a ${\bf \hat{Z}_{t_{s}}}$ gate. (b) Detail of (a).}
 \end{center}
\end{figure}

\section{QUANTUM ERROR CORRECTION}\label{supplementaryerror}

We now consider the situation where the widths of the phase matching and energy conservation conditions are finite and the state obtained corresponds to an ellipse in the JSI plane, as discussed in Section \ref{sectionfour}. As mentioned, in this situation we have an entangled GKP state in time. We can consider that one of the photons, say, idler, plays the role of the ancilla while the signal one is the data qubit in a measurement-based circuit as in Fig. \ref{MBQC}. We will thus perform a measurement in the ancilla (frequency or time measurement) and use the measurement result to correct the data qubit, as in \cite{Steane,Tomthesis}.

\subsection{Correction against temporal shift (MBQC)}

The principle of the MBQC is the following: we prepare an entangled GKP state, noisy in time and frequency (Eq. (\ref{SPDC})), which can be prepared with a SPDC source in an optical cavity as shown in the main text.  Then we perform a time or frequency measurement on the qubit ancilla in a particular basis. Since only the time noise are entangled (see Eq. (\ref{CNOTdispl})), the time measurement provides the information about this displacement see (Fig. \ref{MBQC}) .

We then report the same procedures as in \cite{Steane}, assuming a Dirac distribution for the time and frequency noise and investigate the influence of the time measurement of the ancilla on the time noise of the signal.

We start from a separable state, the data (signal) and the ancilla (idler) are initialized in the frequency $\ket{\overline{+}}_{\omega_{s}}\ket{\overline{+}}_{\omega_{i}}$ state:
\begin{equation}
\ket{\psi}=\ket{\overline{+}}_{\omega_{s}}\ket{\overline{+}}_{\omega_{i}}=\ket{\overline{0}}_{t_{s}}\ket{\overline{0}}_{t_{i}}=\sum_{n,m\in\mathbb{Z}}\ket{nT}\ket{mT},
\end{equation}
with $T=2\pi\tau_{\text{rt}}$. Frequency and time Dirac distribution noises is assumed for both qubits:
\begin{equation}
\ket{\overline{0}}_{t_{s}}\ket{\overline{0}}_{t_{i}}\rightarrow \hat{{\cal{D}}}_{s}(t)\hat{{\cal{D}}}_{i}(t')\hat{D}_{s}(\omega)\hat{D}_{i}(\omega')\ket{\overline{0}}_{t_{s}}\ket{\overline{0}}_{t_{i}},
\end{equation}
then time noises are entangled with the $\hat{C'}$ operation:
\begin{multline}
\hat{C'} {\cal{\hat{D}}}_{s}(t){\cal{\hat{D}}}_{i}(t')\hat{C'} ^{-1}\hat{D}_{s}(\omega)\hat{D}_{i}(\omega')\ket{\overline{0}}_{t_{s}}\ket{\overline{0}}_{t_{i}} \\ ={\cal{\hat{D}}}_{s}(\frac{t+t'}{2}){\cal{\hat{D}}}_{i}(\frac{t-t'}{2})\hat{D}_{s}(\omega)\hat{D}_{i}(\omega')\ket{\overline{0}}_{t_{s}}\ket{\overline{0}}_{t_{i}}\\
=\sum_{n,m} e^{in\omega T}e^{im\omega' T}\ket{nT+\frac{t+t'}{2}}\ket{mT+\frac{t-t'}{2}}.
\end{multline}
We realize a time measurement on the ancilla (the idler), let us consider that the detector clicks at time $\tau$, which can take only the values $\tau=\frac{t-t'}{2}+mT$. The initial state is projected into:
\begin{equation}
\ket{\overline{0}}_{t_{s}}\rightarrow e^{i\omega'(\tau-\frac{t+t'}{2})} {\cal{\hat{D}}}_{s}(\frac{t+t'}{2})\hat{D}_{s}(\omega)\ket{\overline{0}}_{t_{s}}.
\end{equation}

The temporal shift of the data qubit is entirely determined by the noise (shift) of the ancilla. 
The probability of success is given by $\abs{t-t'}<\frac{\pi}{2\overline{\omega}}$, which means the probability to avoid to fall in another $\frac{\pi}{2\overline{\omega}}$ time window.

\subsection{Gaussian distribution of the noise}\label{GaussianMBQC}

Now we consider that the time and frequency noises obey a Gaussian distribution. We thus have the state, as written before Eq. (\ref{writtenbefore}):
{\small{\begin{multline}
\ket{\psi}=[\iint \iint G_{\delta\omega}(\omega)G_{\delta\omega}(\omega')G_{1/\Delta\omega_{-}}(t) G_{1/\Delta\omega_{p}}(t') \\\cross{\cal{\hat{D}}}_{s}(\frac{t+t'}{2}){\cal{\hat{D}}}_{i}(\frac{t-t'}{2})\hat{D}_{s}(\omega)\hat{D}_{i}(\omega')\text{d}t\text{d}t'\text{d}\omega\text{d}\omega']\ket{\overline{0}}_{t_{s}}\ket{\overline{0}}_{t_{i}}.
\end{multline}}}
We then apply the time and frequency displacement operators on the GKP state:
{\small{\begin{multline}
\ket{\psi}=\iint \iint G_{\delta\omega}(\omega)G_{\delta\omega}(\omega')G_{1/\Delta\omega_{-}}(t) G_{1/\Delta\omega_{p}}(t') \\\sum_{n,m\in \mathbb{Z}}e^{in\omega T}e^{im\omega T}\ket{nT+\frac{t+t'}{2}}\ket{mT+\frac{t-t'}{2}}\text{d}t\text{d}t'\text{d}\omega\text{d}\omega' .
\end{multline}}}
The Joint Temporal Amplitude of the state $\bra{t_{s},t_{i}}\ket{\psi}=\text{JTA}(t_{s},t_{i})$ is a circle whose radius is the frequency width $\overline{\omega}$, with elliptical peaks whose half axis are equal to $\Delta\omega_{-}$ and $\Delta\omega_{p}$, see Fig. \ref{JTA}. In the case where $\Delta\omega_{-} \gg \Delta\omega_{p}$, the JTA associated is a periodic (along $t_{-}$) set of lines along $t_{+}$.

We then perform a time measurement on the ancilla, a click is detected at time $\tau$ that can take the value $\tau=mT+\frac{t-t'}{2}$. The new wavefunction $\ket{\psi'}_{\tau}=\bra{\tau}\ket{\psi}$ is, after performing an integration over $t$ and after normalization:
{\small{
\begin{multline}
\ket{\psi'}_{\tau}=\int (\iint \sum_{n,m\in\mathbb{Z}}e^{i\omega nT} \frac{G_{\Delta\omega_{-}}(t'+2(\tau-mT))G_{\Delta\omega_{p}}(t')}{G_{\sqrt{\Delta\omega_{-}^{2}+\Delta\omega_{p}^{2}}}(2(\tau-mT))}\\ \cross G_{\delta\omega}(\omega')G_{\delta\omega}(\omega)e^{i\omega' mT}\ket{(n-m)T+\tau+t')}\text{d}\omega\text{d}t' )\text{d}\omega'.
\end{multline}}}
After the time measurement of the ancilla (idler), the state is projected into a one dimensional GKP state. The time noise distribution of the signal is updated,
\begin{equation}\label{updated}
\frac{G_{\Delta\omega_{-}}(t'+2(\tau-mT))G_{\Delta\omega_{p}}(t')}{G_{\sqrt{\Delta\omega_{-}^{2}+\Delta\omega_{p}^{2}}}(2(\tau-mT))}=G_{\delta}(t'-t_{m}).
\end{equation}
It is a normal distribution with variance $\delta^{2}=\frac{\Delta\omega_{-}^{2}\Delta\omega_{p}^{2}}{\Delta\omega_{-}^{2}+\Delta\omega_{p}^{2}}$ and mean value $t_{m}=\frac{\Delta\omega_{-}^{2}}{\Delta\omega_{-}^{2}+\Delta\omega_{p}^{2}}(\tau+mT)$. The time noise of the data depends on both the noises of the ancilla and the data.

Hence the state can be written as:
\begin{multline}
\ket{\psi'}_{\tau}= \sum_{m\in\mathbb{Z}} \int\iint \text{d}\omega\text{d}t'\text{d}\omega' G_{\delta\omega}(\omega)G_{\delta}(t'-t_{m})G_{\delta\omega}(\omega') \\ \cross e^{i\omega' mT}{\cal{\hat{D}}}_{s}(-mT+\tau+t')\hat{D}_{s}(\omega)\ket{\overline{+}}_{t_{s}}.
\end{multline}
We point out that for time correlated photons with high noise in time variables $\Delta\omega_{-} \gg \Delta\omega_{p}$, the time distribution of the signal only depends on the noise of the idler, since $\delta\sim \Delta\omega_{-}$ and $t_{m}=\tau+mT$. Therefore the analysis is the same than the previous section. We can understand this noise reduction on Fig.\ref{JTA} as follows: when we perform a measurement on the $t_{-}$ axis, the signal is projected into a less noisy state since the updated time distribution of the signal depends on the time distribution of the idler. As a consequence, according to (\ref{updated}), the state becomes periodic along the $t_{+}$ direction, since the time width of each peak becomes $\Delta\omega_{p}$ (instead of $\Delta\omega_{-}$) which is smaller than $2\pi/\overline{\omega}$. 

Note that, if we had considered an anti-correlated initial state, error correction would be effective if the signal photon were detected, instead of the idler one.

\bibliography{biblio2}

\end{document}